\documentclass[fleqn,usenatbib]{mnras}

\usepackage{newtxtext,newtxmath}

\usepackage[T1]{fontenc}

\DeclareRobustCommand{\VAN}[3]{#2}
\let\VANthebibliography\thebibliography
\def\thebibliography{\DeclareRobustCommand{\VAN}[3]{##3}\VANthebibliography}

\usepackage{graphicx}	
\usepackage{amsmath}	

\title{Observational Study of the Atmospheric Gravity Waves in the lower Solar Atmosphere}

\author[Chaurasiya et al.]{
Ravi Chaurasiya,$^{1,2}$\thanks{E-mail: ravi@prl.res.in}
Ankala Raja Bayanna$^{1}$
\\
$^{1}$Udaipur Solar Observatory, Physical Research Laboratory, Udaipur-313001, India \\
$^{2}$ Indian Institute of Technology, Gandhinagar, Gujarat-382355, India}

\date{Accepted XXX. Received YYY; in original form ZZZ}

\pubyear{\the\year{}}

\begin{document}
\label{firstpage}
\pagerange{\pageref{firstpage}--\pageref{lastpage}}
\maketitle

\begin{abstract}

The solar chromosphere exhibits a variety of waves originating from the photosphere and deeper layers, causing oscillations at different heights with distinct frequencies. This study identifies and analyze Atmospheric Gravity Waves (AGWs) and acoustic waves at various height pairs within the solar atmosphere utilising H$\alpha$, Ca {\sc{II}} IR and Fe I 6173 \AA~ imaging spectroscopic observations from Swedish 1m Solar Telescope. We study and compare oscillations by analyzing power maps generated using velocities obtained from the filtergram difference and bisector methods. Our analysis shows a consistent increase in power with height in the solar chromosphere for both methods. In addition to this, our results show that AGWs are detected within or near magnetic flux concentration regions, where spicules are also predominant, exhibiting significant power in the chromosphere. These regions also feature inclined magnetic fields, which might be contributing to the propagation of these low-frequency AGWs in the chromosphere. Examining average power maps at spicule locations reveals significant power at AGWs frequency across different chromospheric heights. We speculate that these AGWs propagate upward along spicular structures and were not previously detected in the studies employing space-time map due to their limited lifetime. This study provides insights into the complex dynamics of solar chromospheric waves influenced by magnetic field, contributing to our understanding of AGWs and acoustic waves propagation across different layers of the solar atmosphere.

\end{abstract}




\section{Introduction} \label{Sec1}

The chromosphere is a thin layer in the Sun's atmosphere. It is found between the cooler surface of the Sun, called the photosphere, and the very hot outer layers, known as the transition region and the corona. 
The chromosphere's density decreases exponentially with height, ranging from about $2 \times 10^{-4}$ kg/m$^3$ at its lower boundary to less than $1.6 \times 10^{-11}$ kg/m$^3$ at its upper boundary \citep{2008A&A...489L..57K}. In contrast, its temperature varies irregularly, initially dropping to 3800 K \citep{2003ASPC..286..419A} before rising to over 15000 K in the upper layers. While much of the studies over the past decades has concentrated in understanding the  coronal heating, the chromosphere presents a significant opportunities for advancing our solar as well as astrophysical understanding.
Even though the chromosphere is a thin layer, it plays a crucial role as the interface between the cooler photosphere and the multi-million degree heated corona. Although the chromosphere is only a bit hotter than the photosphere, it is much denser than the corona. Because of this, it needs about twice as much energy to balance out the energy it loses through radiation (\cite{1977ARA&A..15..363W,1989ApJ...336.1089A}). 
Therefore, the chromosphere is important for understanding how energy transports between the photosphere and the corona.

The chromosphere is most readily observed in spectral lines such as H$\alpha$, Ca II 854.2 nm, Ca II H and K, and He I 1083 nm. 
Observations in these spectral lines reveal that the chromosphere is dominated by jets, such as spicules (\cite{secchi1877soleil,2004Natur.430..536D}), and various types of waves, both acoustic and non-acoustic, which lead to rapid dynamic changes within timescales ranging from a few seconds to minutes. Spicules are thin, hair-like structures seen at the solar limb in chromospheric spectral lines. They move upward and downward in the solar atmosphere and exhibit heating to transition region and corona temperatures (\cite{2011Sci...331...55D,2014ApJ...792L..15P,2015ApJ...799L...3R,2016ApJ...820..124H,2019Sci...366..890S,2024ApJ...970..179C}). Recently, \cite{2024ApJ...970..179C} suggested that spicules might also contribute to chromospheric heating, in addition to their role in heating the transition region and corona. Additionally, studies focused on waves
has gained a lot of attention because they can carry mechanical energy and help us understand the Sun's physical paramters through a method called seismology (\cite{2009SSRv..149..355Z,2012RSPTA.370.3193D,2015SSRv..190..103J,2019ARA&A..57..189C,2023LRSP...20....1J}).

Acoustic waves, or p-modes, originate in the solar interior's convection zone and are generally confined as they are reflected back due to a sharp drop in sound speed near the sun's surface. However, these waves can propagate easily from the surface into the atmosphere if their periods are shorter than 3.2 minutes (or the frequency greater than 5.2 mHz), known as the acoustic cut-off period (frequency) of the photosphere (\cite{1979ApJ...231..570L,1992ApJ...397L..59C}). While longer-period waves typically do not reach higher atmosphere, there is still a substantial observational evidence of oscillations in the chromosphere around network magnetic elements (\cite{2007A&A...461L...1V,2014MNRAS.443.1267B,2023LRSP...20....1J}). Studies suggests that the strong magnetic fields can change the radiative relaxation time, potentially increasing the cut-off period significantly (\cite{2006ApJ...640.1153C,2008ApJ...676L..85K}).
 On the other hand, the inclination of the magnetic field also plays a crucial role in the propagation of long-period waves (\cite{2004Natur.430..536D,2006RSPTA.364..395C,2006ApJ...648L.151J,2013ApJ...779..168J,2014A&A...567A..62K}). 
The 2D map of power distribution across different heights in the solar atmosphere indicates that for the 3-minute oscillation, while power is enhanced in the upper photosphere (power halos), it is suppressed in the chromosphere (magnetic shadows), particularly in and around the network regions. Studies suggest that the interaction between acoustic waves and magnetic fields leads to the formation of these magnetic shadows and power halos (\cite{1992ApJ...393..782T,2001ApJ...554..424J,2001ApJ...561..420M,2003A&A...405..769M,2007A&A...471..961M,2010A&A...524A..12K,2013SoPh..287..107R,2016ApJ...828...23S,2023LRSP...20....1J}).

\begin{figure*}
	\centering
	\includegraphics[width=150mm]{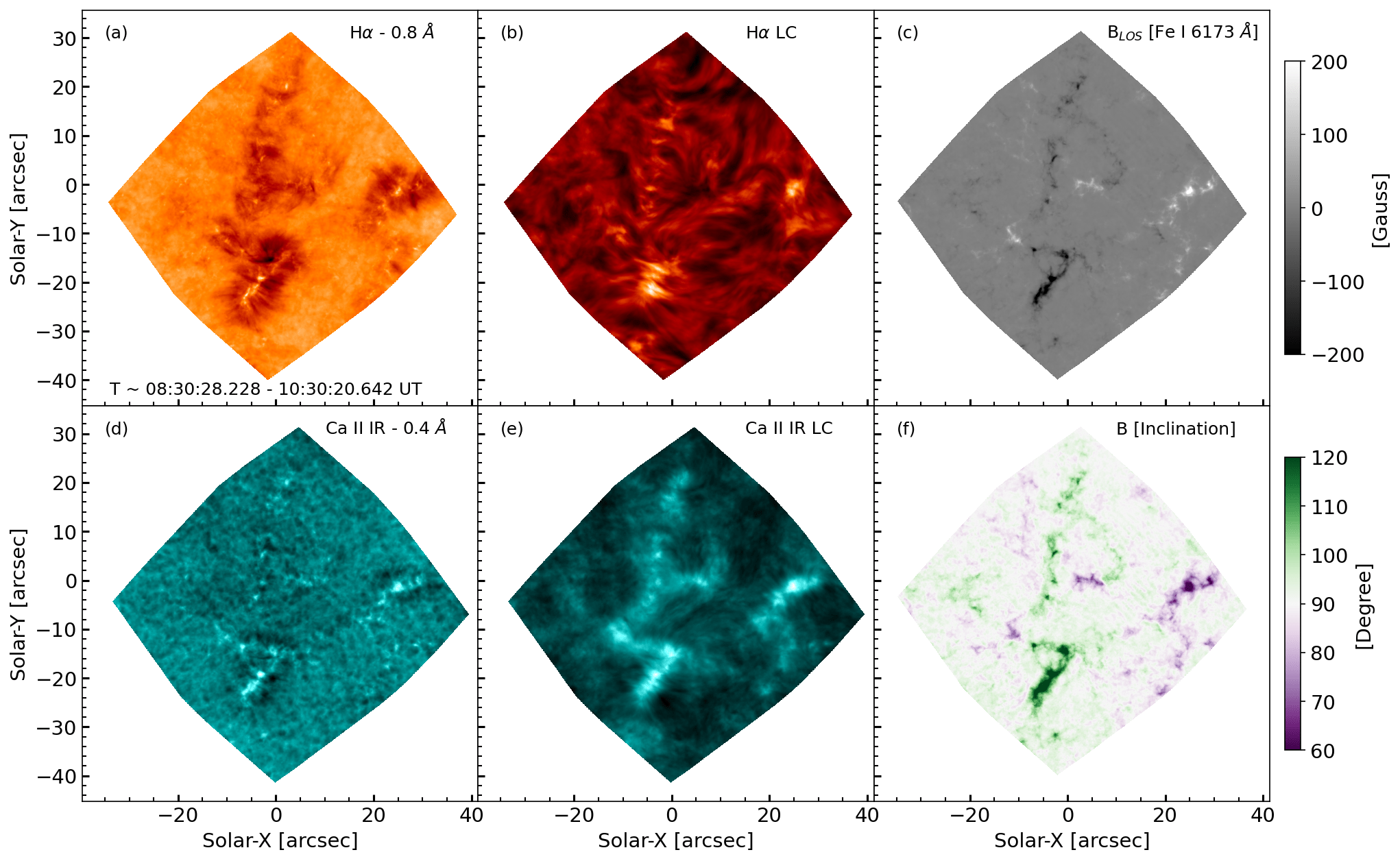}
	\caption{Panels $\textbf{(a)}$ and $\textbf{(b)}$ show the average blue and line centre images of H$\alpha$, while panels {\bf (d)} and $\textbf{(e)}$ display the average blue and line centre images of the Ca II IR spectral line, respectively. Panels \textbf{(c)} and \textbf{(f)} depict the photospheric LOS magnetic field and its inclination, derived from the observations using Milne-Eddington inversion code \protect\citep{2019A&A...631A.153D}. Here, the inclination is measured with respect to the normal to the surface such that 90$^\circ$ represent the field parallel to the surface. The dark region surrounding the magnetic flux concentration appearing in the blue wing images of H$\alpha$ and Ca II IR are the region dominated by the spicules.}
	\label{Fig1}
\end{figure*}

In addition to acoustic waves, the solar atmosphere also contains low-frequency waves that utilize buoyancy as a restoring force and exhibit unique polarization properties. These waves are called Atmospheric Gravity Waves (AGWs) and are generated by turbulent subsurface convection overshooting or locally penetrating a stably stratified medium \citep{1967IAUS...28..429L,1981ApJ...249..349M,1982ApJ...263..386M,2008A&A...489L..57K,2019ApJ...884L...8J}. Similar to \cite{2023ApJ...952...58V}, we also use the term AGWs to refer propagating gravity waves in the solar atmosphere. It’s quite challenging to directly observe and track these waves because they are thought to become very weak in the convection zone. Their amplitude at the surface is very small, and they have low temporal frequencies and transverse movement ( \cite{2018SoPh..293...95S,2021RSPTA.37900178C}). However, understanding these waves is important because it helps us to understand about their own properties and can also be used to diagnose the magnetic field in the chromosphere, since these waves are greatly influenced by magnetic fields. AGWs exhibit a distinctive negative phase difference while transporting energy upward from the photosphere to higher atmospheric layers \citep{1978cup..book.....L}, with their group and phase velocity being perpendicular to each other. This behavior can be clearly identified by estimating the phase spectra obtained from velocity/intensity measurements at two different heights.

The propagation characteristics of AGWs in the solar atmosphere have been extensively studied through velocity-velocity (V--V) and intensity-intensity (I--I) observation pairs from various ground- and space-based facilities, as well as 3D realistic numerical simulations (\cite{1967IAUS...28..429L,2008ApJ...681L.125S,2011A&A...532A.111K,2014SoPh..289.3457N,2017ApJ...835..148V,2019ApJ...872..166V,2020A&A...633A.140V,2021RSPTA.37900177V,2023AdSpR..72.1898K,2023ApJ...952...58V}). 
Utilizing V--V observation pairs and comparing with the 3D numerical simulations, \cite{2008ApJ...681L.125S} detected AGWs, estimating that the energy flux carried by these AGWs are comparable to the radiative losses of the chromosphere. Later on, \cite{2011A&A...532A.111K} using observations of Fe I 5576 \AA~and Fe I 5434 \AA~lines, studied AGWs in the Quiet Sun (QS) concluding that gravity waves also contribute to chromospheric heating hence reinforcing the result made by \cite{2008ApJ...681L.125S}. Studies by \cite{1981ApJ...249..349M,1982ApJ...263..386M} suggested that gravity waves can reach high in the chromosphere (900–1600 km) depending upon the energy flux of these AGWs.

These AGWs are also influenced by the magnetic field, which can be utilized for estimating the average magnetic field strength (\cite{2011MNRAS.417.1162N,2017ApJ...835..148V,2019ApJ...872..166V,2020A&A...633A.140V,2021RSPTA.37900177V}). Recent realistic 3D numerical simulations of the solar atmosphere, using the CO${^5}$BOLD code \citep{2012JCoPh.231..919F}, have demonstrated that AGWs are generated abundantly and propagate independent of the field strength in the low photosphere (\cite{2017ApJ...835..148V,2019ApJ...872..166V,2020A&A...633A.140V,2021RSPTA.37900177V}). 
\cite{2017ApJ...835..148V} found that AGWs are absent or partially bounce back into the lower layers in the presence of vertical magnetic fields. However, horizontal magnetic fields would allow these waves to reach chromospheric heights \citep{2021RSPTA.37900177V}, where they encounter wave-breaking heights as discussed by \cite{1981ApJ...249..349M,1982ApJ...263..386M}.

Utilizing multi-wavelength data from the Interferometric BIdimensional Spectrometer (IBIS; \cite{2006SoPh..236..415C}), Michelson Doppler Imager (MDI; \cite{1995SoPh..162..129S}), and Solar Dynamics Observatory (SDO; \cite{2012SoPh..275....3P}), \cite{2008ApJ...681L.125S,2023AdSpR..72.1898K} have provided observational evidence of the suppression of AGWs at locations of strong magnetic flux. More recently, \cite{2023ApJ...952...58V} found that, at least near the quiet-Sun disk center, the lower photospheric magnetic field does not significantly affect the generation and propagation of AGWs. 
While emphasis has been given to the effects of vertical and horizontal magnetic fields on AGW propagation, the impact of inclined magnetic fields has not yet been explored observationally.

Here, we study the propagation of AGWs from the photosphere to the high chromosphere and their association with the magnetic field inclination and their possible association with spicules. In this regard, we utilize high-resolution observations in H$\alpha$, Ca II IR, and Fe I 6173 Å to investigate the oscillation properties at different solar atmospheric layers by constructing power maps at various heights. We then analyze the propagation of acoustic waves and AGWs at multiple heights by estimating phase difference and coherency spectra, along with their association with the magnetic field and its inclination at the photosphere.

The rest of the paper is structured as follows: Section \ref{Sec2} describes the observations and the spectropolarimetric inversion of the photospheric spectral line. Section \ref{Sec3} presents the results obtained, including the comparison of the power maps and the propagation of waves at multiple heights in the solar atmosphere through the estimation of phase difference and coherency spectra. The results are discussed and summarized in Section \ref{Sec4}, before concluding the paper in Section \ref{Sec5}.

\section{Observations} \label{Sec2}

We analyzed a dataset consisting of QS observations at the disk center, recorded on 24 April 2019, from approximately 08:30 UT to 10:30 UT by the Swedish 1-m Solar Telescope (SST; \cite{2003SPIE.4853..341S}). The field-of-view (FoV) also includes weak magnetic flux concentration as shown in Figure \ref{Fig1}. This datasets consist of imaging spectroscopic observations of H$\alpha$ and Ca II IR for chromospheric observation, and Fe I 6173 Å for photospheric observation, obtained using the CRisp Imaging Spectropolarimeter (CRISP; \cite{2008ApJ...689L..69S}). These observations spanned nearly 2 hours, with a temporal cadence of 28.185 seconds. The FoV was approximately 60" $\times$ 60" with a plate scale of 0.0591" per pixel. The H$\alpha$, Ca II IR, and Fe I 6173 Å lines were sampled at 25, 20, and 15 wavelength positions, respectively, with spectral coverage of $\pm$ 1.2 Å, $\pm$ 0.8 Å, and $\pm$ 0.245 Å around the line center. The effects of atmospheric seeing in the data were corrected using the SST adaptive optics system (\cite{2019A&A...626A..55S}) and then further improved with the Multi-Object Multi-Frame Blind Deconvolution (MOMFBD; \cite{2005SoPh..228..191V}) image restoration method. This dataset is publicly available from the SST archive\footnote{https://dubshen.astro.su.se/sstarchive/search}, provided by the SST Team. All three datasets were aligned by cross-correlating the nearly simultaneous wing filtergrams of each observation, which are dominated by photospheric features. To infer the photospheric line-of-sight (LOS) magnetic field and its inclination along with the LOS velocity, we used the Milne-Eddington (ME) inversion code\footnote{https://github.com/jaimedelacruz/pyMilne}, developed by \cite{2019A&A...631A.153D}, which utilizes the full Stokes profiles of Fe I 6173 Å for the inversion process. The field is measured relative to the normal to the surface, where the field perpendicular to the surface is defined as \(0^\circ\), while the field parallel to the surface (horizontal field) is defined as \(90^\circ\). The noise level in the average LOS magnetic field (B$_{LOS}$) was determined by taking the standard deviation of a small region (100 $\times$ 100 pixels) located away from the magnetic flux concentration area. The noise level was found to be less than 2 G. The  blue wing and line center filtergrams of H$\alpha$ and Ca II IR, along with the LOS magnetic field and inclination, averaged over the observation duration are shown in Figure \ref{Fig1}.

\section{Power distribution at various heights in the solar atmosphere}  \label{Sec3}

\subsection{Determination of time-varying signal for power estimation}

\begin{figure}
	\centering
	\includegraphics[width=85mm]{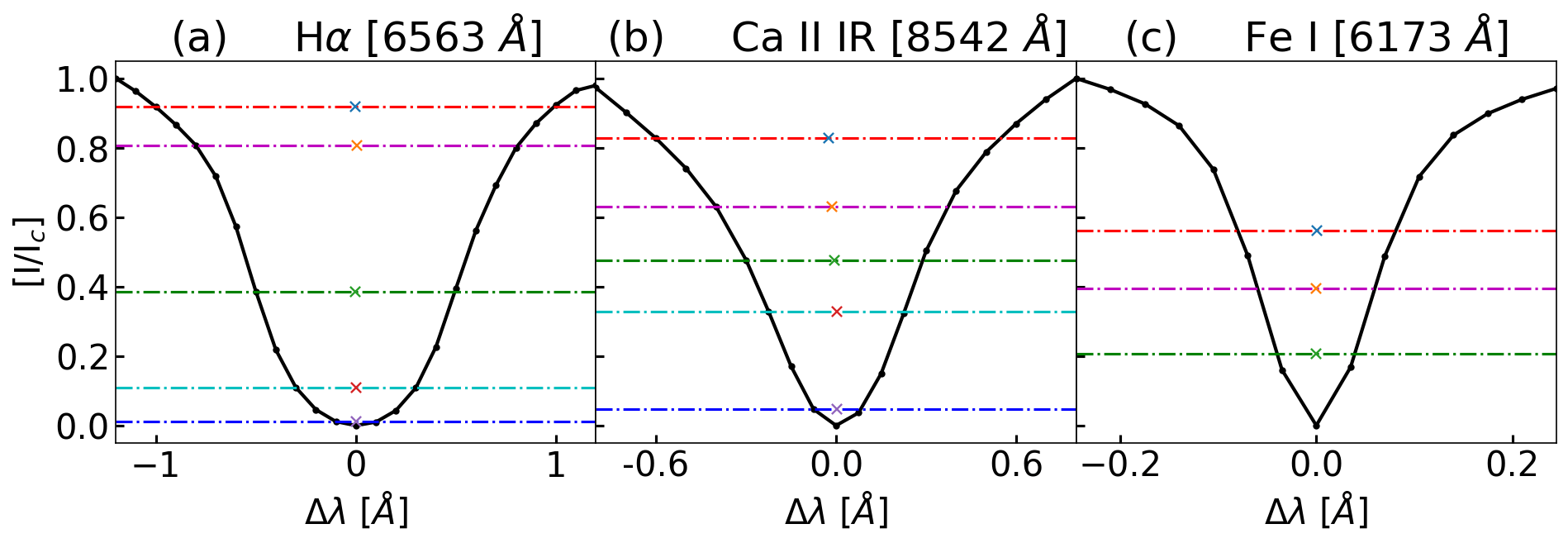}
	\caption{Panels \textbf{(a)}, \textbf{(b)}, and \textbf{(c)} represent the normalized average spectra of H$\alpha$, Ca II IR, and Fe I 6173 \AA~respectively. The coloured horizontal lines indicate the intensity levels at which bisector velocities and filtergram differences were calculated for power estimation. These intensity levels correspond to H$\alpha$ $\pm$ 1.0, $\pm$ 0.8, $\pm$ 0.5, $\pm$ 0.3, $\pm$ 0.1 \AA~, Ca II IR $\pm$ 0.6, $\pm$ 0.4, $\pm$ 0.3, $\pm$ 0.225, $\pm$ 0.075 \AA~, and Fe I 6173 \AA~ $\pm$ 0.08, $\pm$ 0.06, $\pm$ 0.04 \AA. The cross marks '\textbf{x}' indicate the bisector offset/velocity corresponding to each respective intensity level and certain solar atmosphere heights.}
	\label{Fig2}
\end{figure}

H$\alpha$ and Ca II IR spectral lines are renowned for chromospheric observations and Fe I 6173 \AA~ line for photospheric observations. However, each of these spectra captures information from various heights in the solar atmosphere because different parts of the spectrum form at different heights. Specifically, the filtergrams at the line centers of the H$\alpha$ and Ca II IR spectra represent the higher and mid-chromosphere, respectively. Conversely, the filtergrams at the far wings of these lines are formed near the photosphere, while those at the inner wings sample higher layers above the photosphere (\cite{2006A&A...449.1209L, 2012ApJ...749..136L, 2014ApJ...781..126O}). Similarly, for the Fe I 6173 \AA~ spectral line, the filtergrams at the wings are formed in the lower photosphere, whereas the filtergram at the line center corresponds to the upper photospheric layer.

To study and compare oscillations in velocity at various heights in the solar atmosphere, we use filtergram difference and bisector methods to estimate the velocity. Both these methods estimates Doppler shifts, thus the Doppler velocity.

$\textbf{1. Filtergram Difference Method:}$ This involves calculating the normalised difference between filtergrams obtained at the same wavelength offsets in the blue and red wings of the spectral lines \citep{2004ApJ...604..906R}. This method provides information about oscillations based on intensity variations at specific wavelength offsets from the line center. To calculate velocities from filtergram differences, we multiply the normalised intensity difference by a scaling factor as shown in equation \ref{eq:velocity}.

The Doppler shift, \(\Delta \lambda\), is given by the following expression:

\begin{equation}
	\Delta \lambda = k \frac{I_b - I_r}{I_b + I_r}
	\label{eq:velocity}
\end{equation}

where, \(k\) is the scaling factor, and \(I_b\) and \(I_r\) represent the blue and red wing intensities of a spectral line profile, respectively. This scaling factor is determined by shifting the average spectral line profile with artificial velocities.
It should be noted that the value of \(k\) varies with the wavelength offset for which filtergram difference to be obtained. To obtain the  multi-height velocities, we obtain \(k\) values for different wavelength offsets across the given spectral line profile.

$\textbf{2. Bisector Method:}$ In this method, Doppler shift is obtained from the mean of the red wing and blue wing wavelengths, for a given intensity level. By choosing different intensity levels across the line profile, we estimate the multi-height Doppler shifts. This method account for the spectral line assymetries. 

From now onwards, we refer to filtergram difference velocity and bisector velocity for the Doppler shift/velocity obtained from the filtergram difference and bisector methods respectively.

Panels (a) and (b) of Figure \ref{Fig2} plot the intensity levels at which the filtergram difference and bisector velocities are estimated over the average spectral lines of H$\alpha$ and Ca II IR, respectively. The cross marks '\textbf{x}' indicate the bisector offset/velocity at each height in the solar atmosphere. Panel (c) of Figure \ref{Fig2} shows the intensity levels for which only bisector velocities are estimated for the Fe I 6173 \AA~ spectral line. These multi-height velocites are then averaged to get the average bisector velocity of the photosphere.

\begin{figure}
	\centering
	\includegraphics[width=85mm]{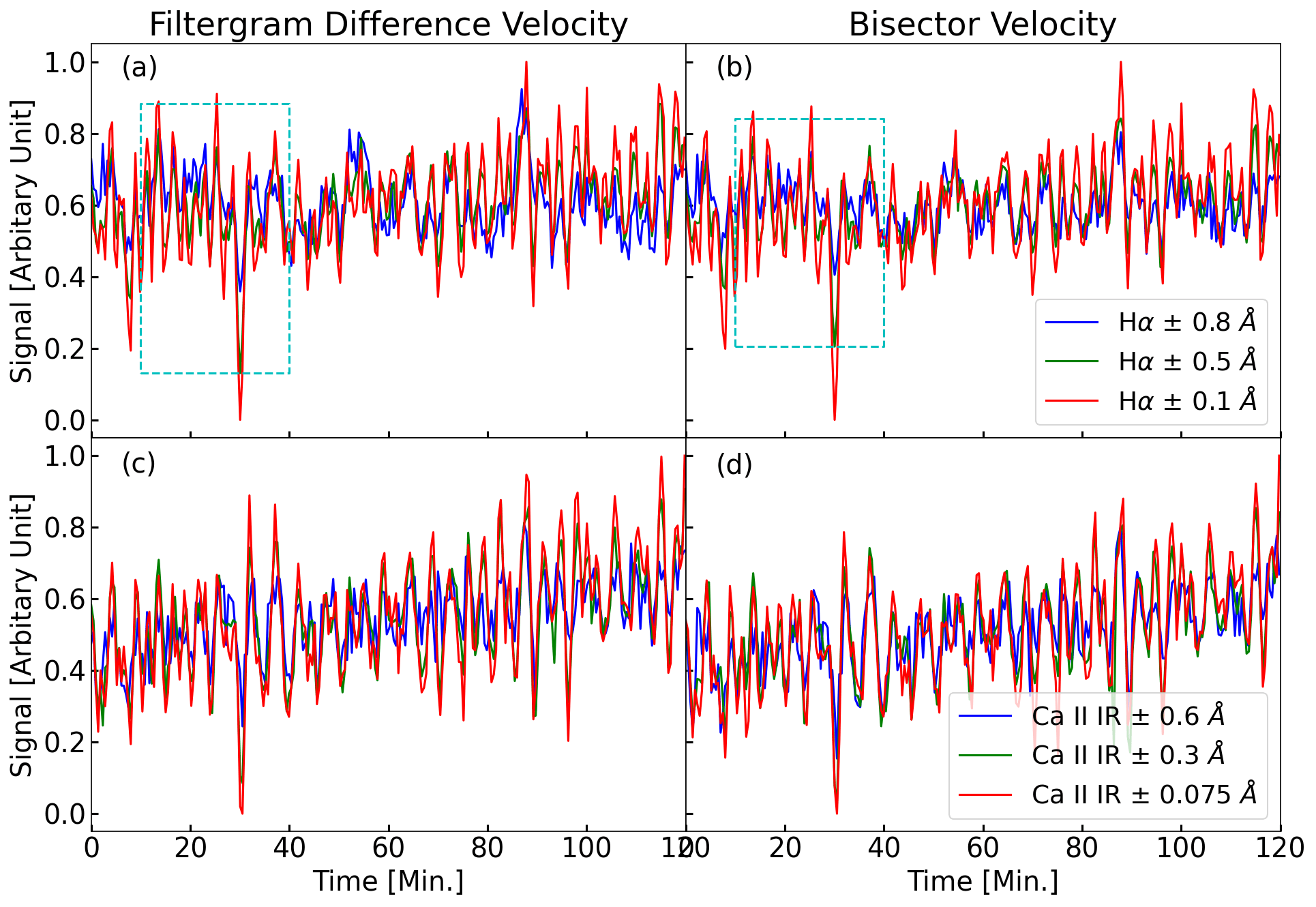}
	\includegraphics[width=85mm]{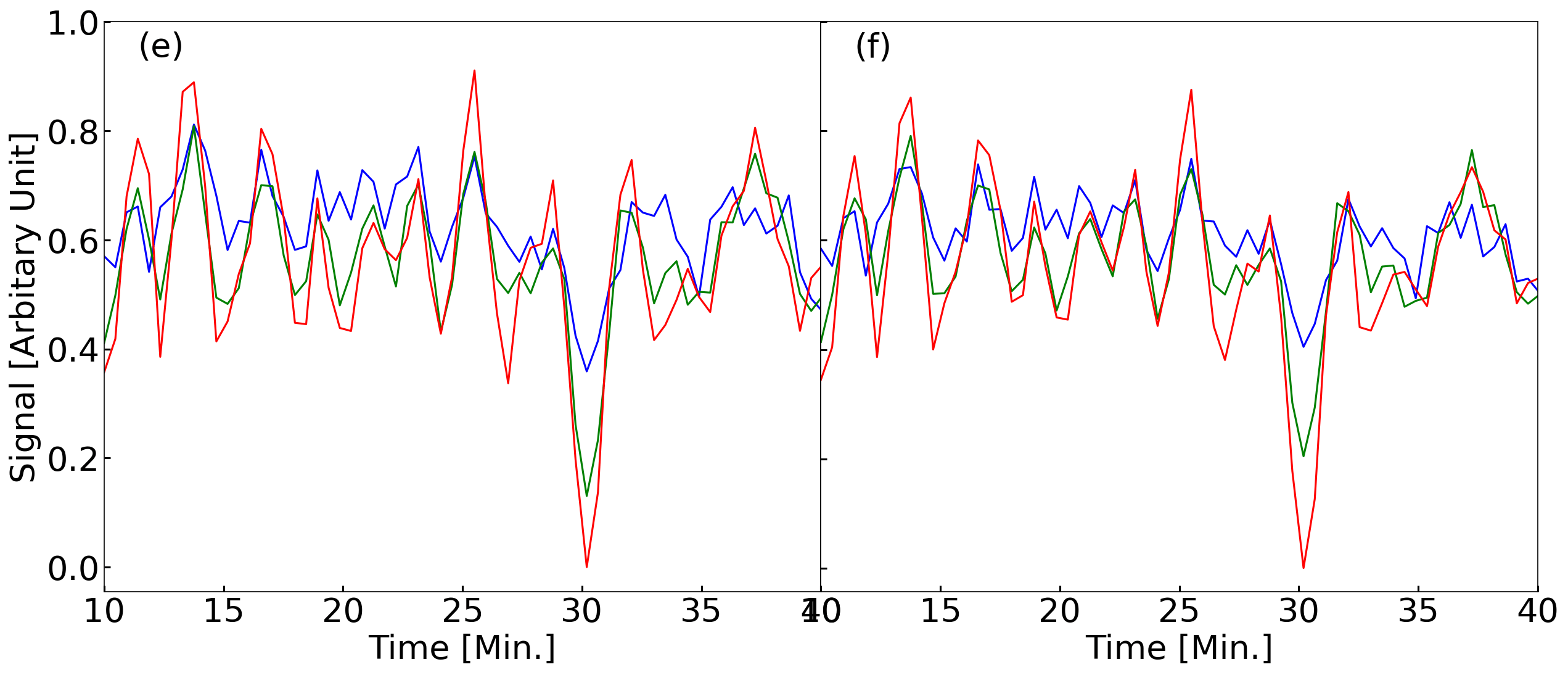}
	\caption{Panel \textbf{(a)} -- \textbf{(d)} shows the average time-varying signal originating from different heights in the solar atmosphere. While Panels \textbf{(a)} and \textbf{(b)} illustrate the filtergram difference and bisector velocity signal of H$\alpha$, panels\textbf{ (c)} and \textbf{(d)} illustrate Ca II IR, respectively. Panels \textbf{(e)} and \textbf{(f)} represent the zoomed-in views (cyan rectangle in Panels \textbf{(a)} and \textbf{(b)}) of 30 minutes signal. The comparison of the amplitude of the average time-varying signal obtained from filtergram difference and bisector velocities suggests that the upper atmosphere dominates in both the method used. This points towards the indication that the same would be true for power as power is proportional to the strength of the signal.}
	\label{Fig3}
\end{figure}

Before obtaining the power maps, we first examine the average variation in the amplitude of the signal across the entire observing period, as illustrated in Figure \ref{Fig3}. This figure shows the temporal variation of the Doppler shifts, averaged over the entire FoV for different heights in the solar atmosphere. Estimating these average signals helps identify the atmospheric layer where the power is most dominant, as the strength of the time-varying signal is directly proportional to wave power. Although we have estimated filtergram differences and bisector velocities at five different heights in the solar atmosphere, for clarity, Figure \ref{Fig3} presents the average signals at only three specific heights for both chromospheric spectra: H$\alpha$ $\pm$ 0.8, $\pm$ 0.5, $\pm$ 0.1 \AA~ and Ca II IR $\pm$ 0.6, $\pm$ 0.3, $\pm$ 0.075 \AA. By comparing the signals from filtergram differences and bisector velocities in H$\alpha$ and Ca II IR, it is evident that the amplitude of the signal increases as one goes high in the solar chromosphere.

\subsection{Comparison of power obtained from filtergram difference and bisector velocity}\label{ssec3}

\begin{figure*}
	\centering
	\includegraphics[width=180mm]{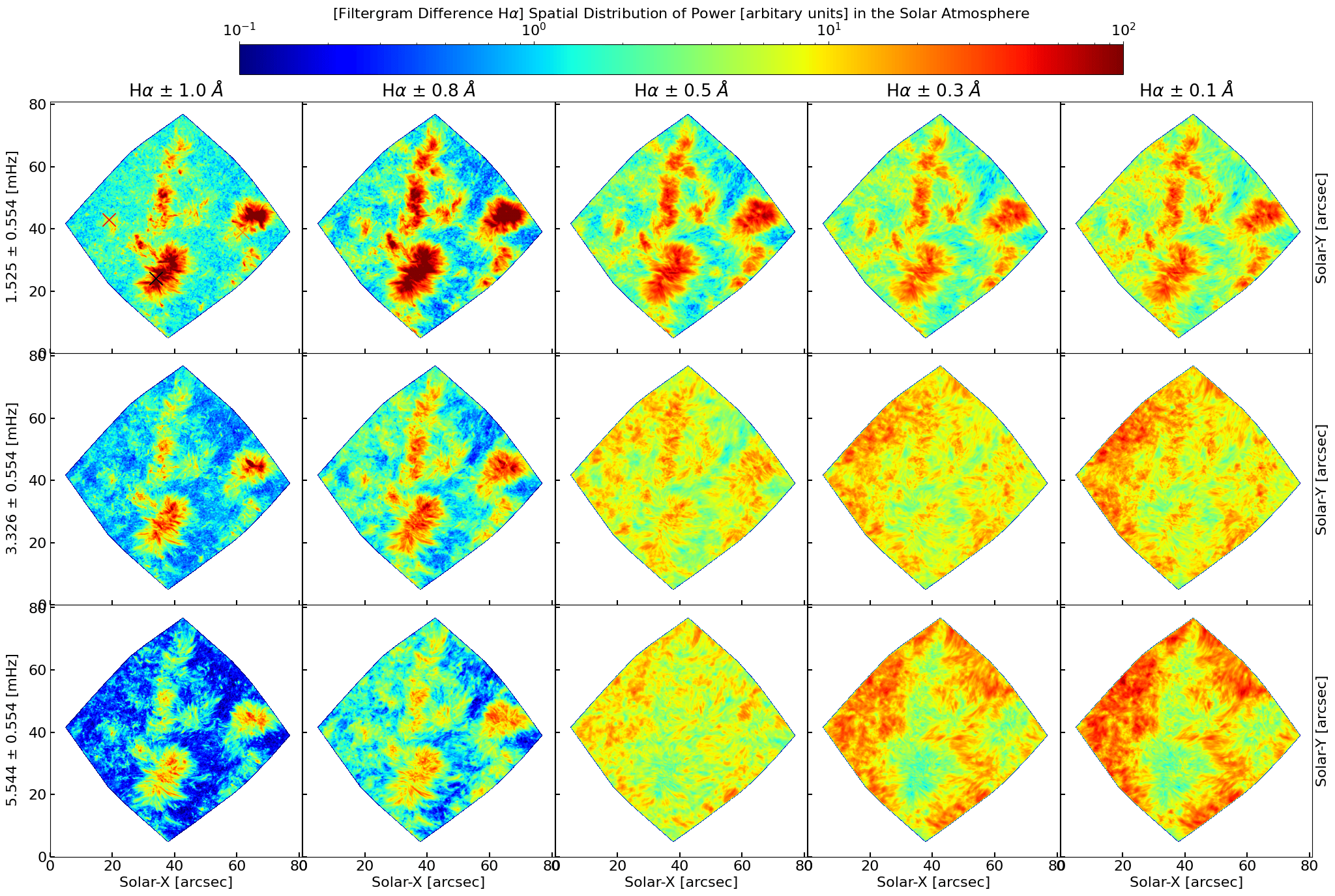}
	\caption{Power distribution from H$\alpha$ filtergram difference: The above panel shows 5 columns (representing 5 different heights in the solar atmosphere in increasing order from left to right, starting from the photosphere) and 3 rows (representing 3 frequencies in increasing order from top to bottom). These rows show power distribution maps corresponding to frequencies of 1.525, 3.326, and 5.544 mHz, obtained by averaging the power in a 1.1 mHz frequency band around each frequency. The power spectrum corresponding to the three different locations ( \textbf{X} marks, as shown in the upper right panel) is depicted in the upper panel of Figure \ref{Fig6}.}
	\label{Fig4}
\end{figure*}

\begin{figure*}
	\centering
	\includegraphics[width=180mm]{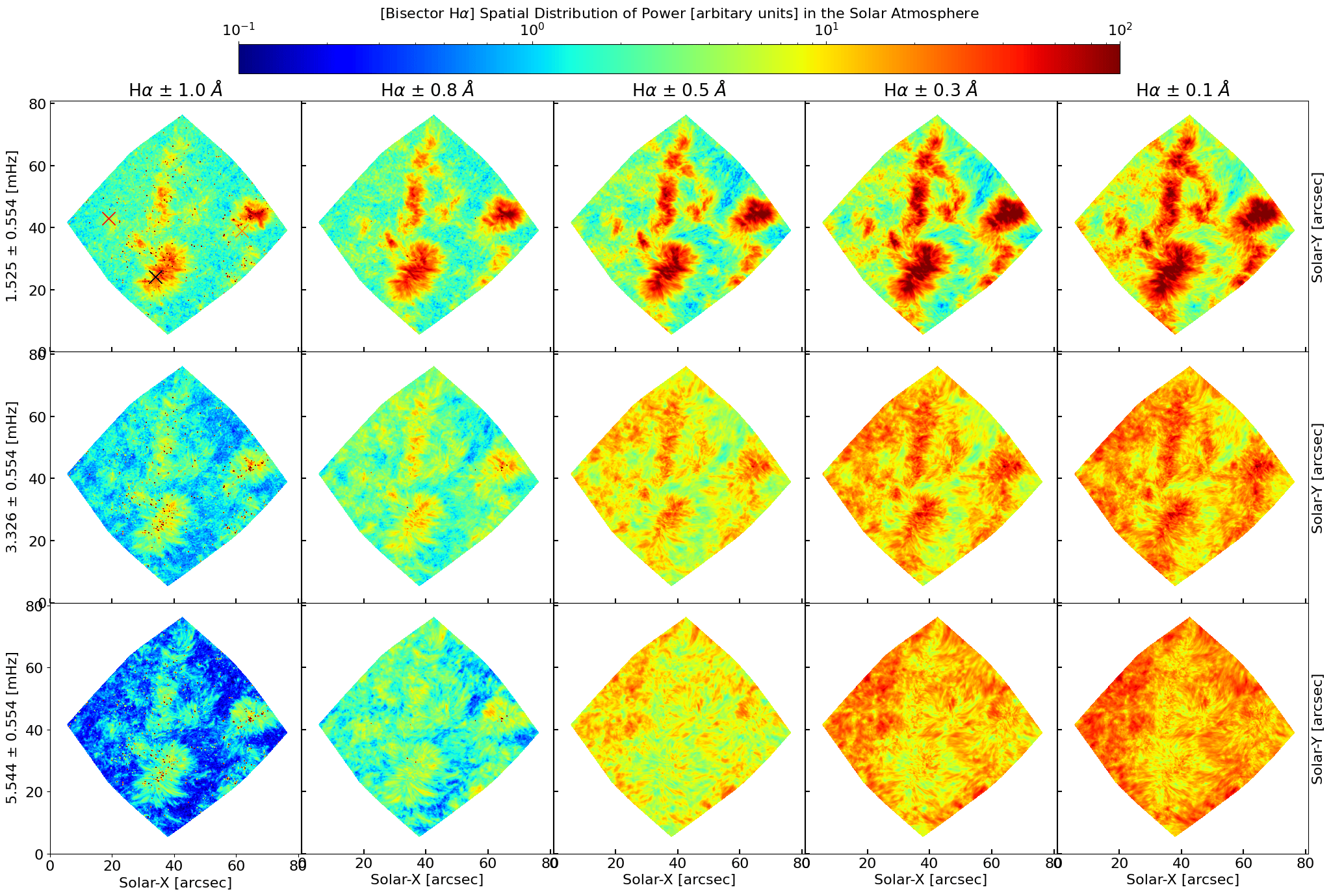}
	\caption{Same as Figure \ref{Fig3} but for the bisector velocity. The power spectrum corresponding to the three different locations ( \textbf{X} marks, as shown in the upper right panel) is depicted in the lower panel of Figure \ref{Fig6}.}
	\label{Fig5}
\end{figure*}

We apply Fourier transforms to the mean substracted filtergram difference velocities and bisector velocities for each time series signal corresponding to various heights in the solar atmosphere to estimate power distributions. The sets of power maps obtained from filtergram differences and bisector velocities for H$\alpha$ and Ca II IR spectra at different heights (refer to Figure \ref{Fig2}) are depicted in Figures \ref{Fig4}, \ref{Fig5} and in Figures \ref{FigA1} and \ref{FigA2}, respectively. Each figure consists of 5 columns and 3 rows, representing the power distribution maps corresponding to 1.5, 3.3, and 5.5 mHz frequencies, obtained by averaging the power in a 1.1 mHz frequency band around each frequency.  

\begin{figure}
	\centering
	\includegraphics[width=85mm]{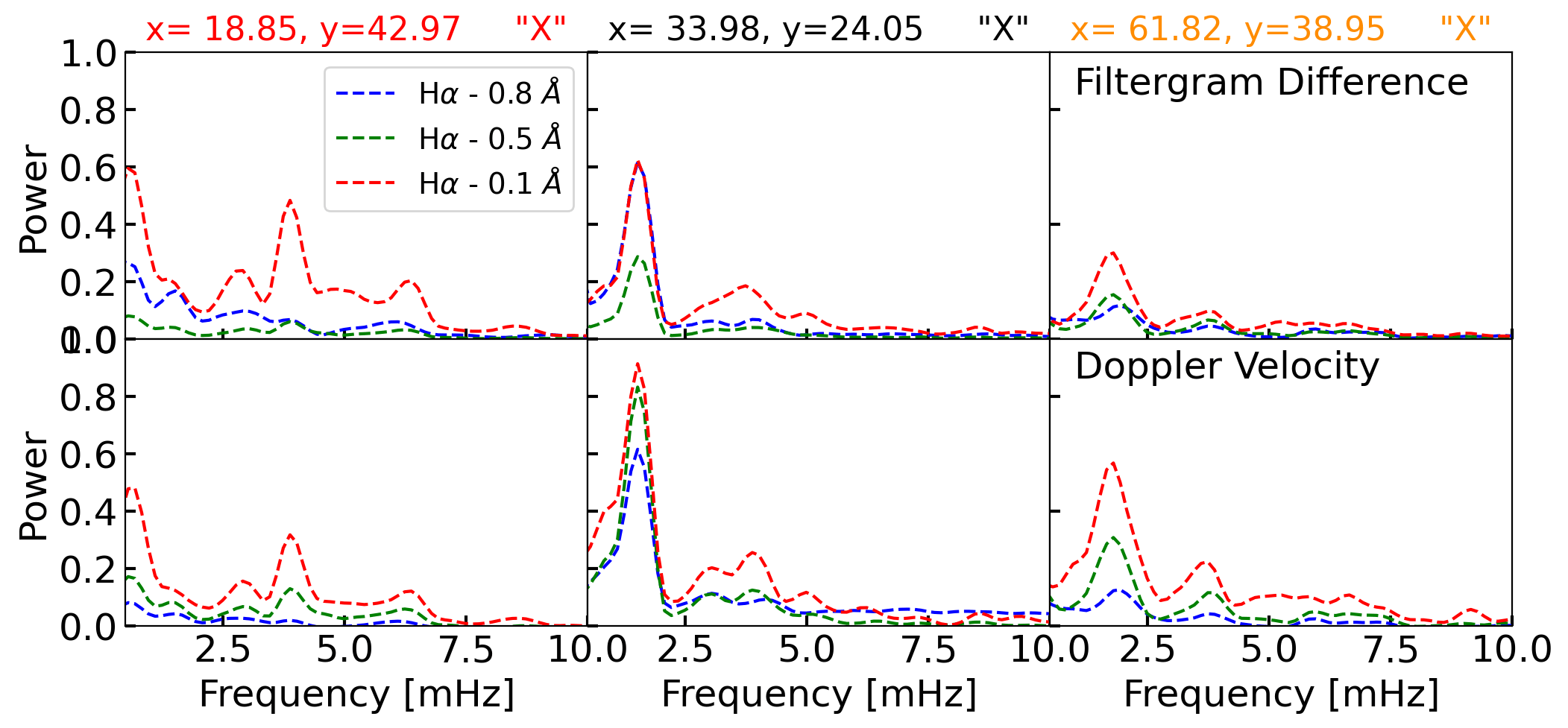}
	\caption{The above plots show three power spectra (corresponding to three spatial locations marked by colored "\textbf{X}" in Figure \ref{Fig4} and Figure \ref{Fig5} from left to right) obtained at three heights in the solar atmosphere using filtergram difference (upper panel) and bisector velocity (lower panel) as a time-varying signal. Comparing the power spectra indicates that the dominant power corresponding to certain frequencies is consistent for both methods used to estimate the power spectrum.}
	\label{Fig6}
\end{figure}

Comparing the power distribution at multiple heights and at different frequency regimes suggests:

\begin{itemize}
	
	\item In the low frequency regime, power in or near magnetic regions is consistently dominant than the non-magnetic regions. 
	
	\item In the high-frequency regime, while power is enhanced in and around magnetic regions compared to the quiet Sun in the upper photosphere or lower chromosphere, it is suppressed in the higher chromosphere. These effects are referred to as power halos and magnetic shadows, respectively. It seems that the transition from power halos to magnetic shadows occurs after the atmospheric height corresponding to H$\alpha$$\pm$0.8 Å (or Ca II IR $\pm$0.4 Å) in the solar atmosphere.

	\item A similar behavior is observed in the Ca II IR observations shown in Figures \ref{FigA1} and \ref{FigA2}.
	
	\item A closer look at the power maps suggest that the dominant power corresponding to certain frequencies remains consistent for both methods at a given location. To verify this, power spectra at three spatial locations and at three heights in the solar atmosphere, averaged over a 3 $\times$ 3 - pixel neighbourhood around the marked pixel is depicted in Figure \ref{Fig6}. The same pattern holds true for Ca II IR, as presented in  Figure \ref{FigA3}.
	
\end{itemize}

\begin{figure}
	\centering
	\includegraphics[width=55mm]{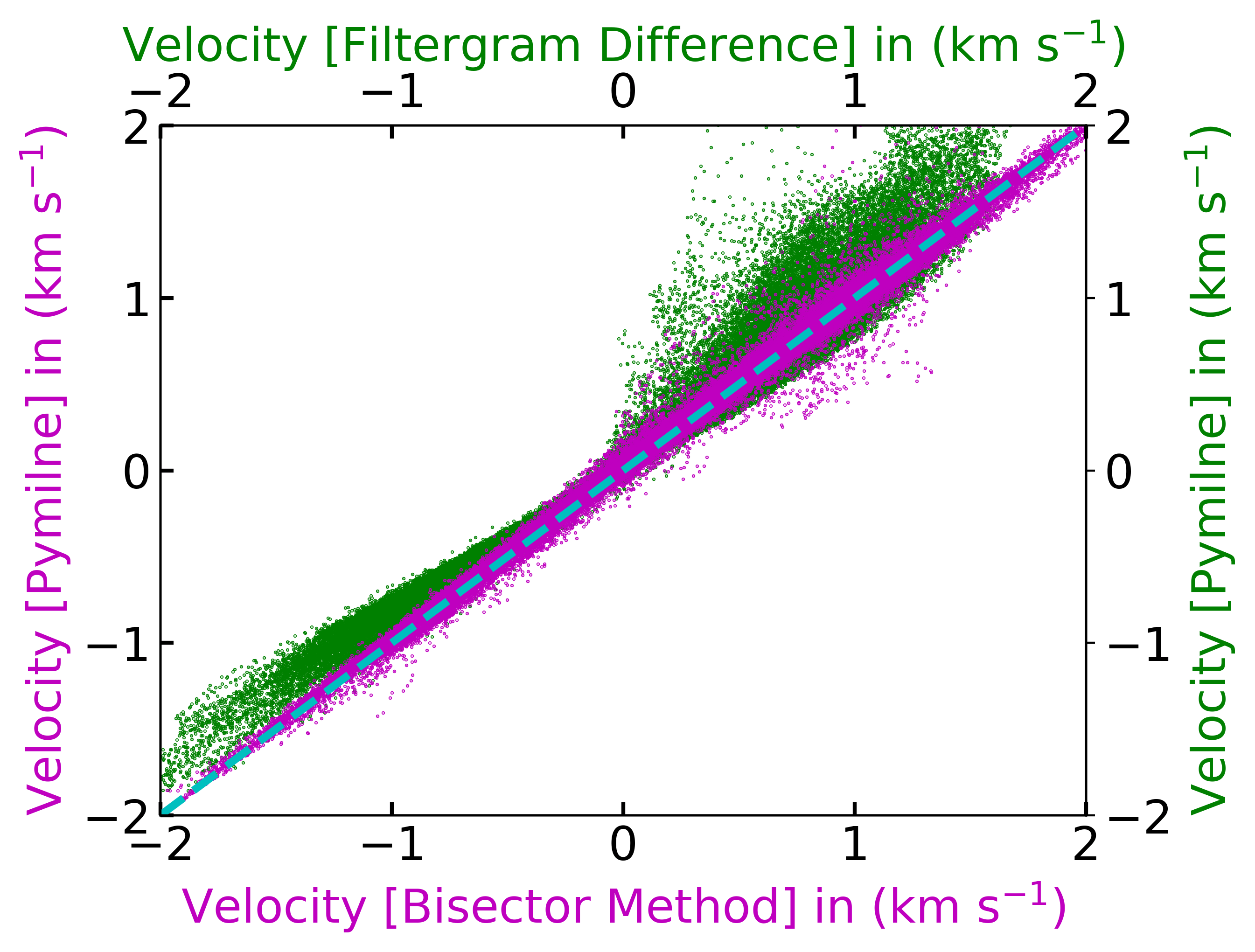}
	\caption{Comparison of velocity in the photosphere utilising Fe I 6173 \AA~ observations obtained from bisector method and ME inversion (magenta color) and filtergram difference and ME inversion (green color). The velocities shown here are averaged over the multiple heights in case of bisector and filtergram difference. The dashed straight line in cyan color represents y=x.}
	\label{FigA4}
\end{figure}

It is to be noted that velocities obtained from bisector and filtergram difference methods do not correlate very well. In order to verify this, we obtained Doppler velocity from the ME inversion of the Fe I 6173 \AA~ observations and compared with velocities obtained from these two methods in the photosphere. It is evident from Figure \ref{FigA4} that the velcoity obtained from ME inversion correlates well with that of the bisector method but shows considerable spread at larger velocities in the case of filtergram difference. However, the power maps derived from velocities obtained from filtergram difference and bisector methods show similar results for both the chromospheric spectral lines, as discussed earlier. Hence, we will proceed with further analysis using only the power maps derived from bisector velocity as the time-varying signal for both the photospheric and chromospheric spectral lines.

\subsection{Identification of the location of propagating waves in the solar atmosphere}

Although power maps at various heights indicate that the chromosphere predominantly exhibits oscillations at frequencies of 1.5, 3.3, and 5.5 mHz, all these waves do not propagate from the lower to the upper solar atmosphere across all locations. Wave propagation to higher layers is influenced by several factors, including power, frequency, magnetic field strength, and magnetic field inclination to name a few.
To analyze wave propagation, we estimate the phase difference (the phase at the upper layer minus the phase at the lower layer) and coherency spectra (normalized cross spectral density) between two specific heights using bisector velocity as the time-varying signal. A non-zero phase difference indicates the presence of propagating waves, whereas a zero phase difference suggests standing waves or non propagating waves. High coherency values close to 1 imply strong linear dependence between waves at the two heights, while low coherency values near zero suggest weak linear dependence. In our analysis, we identify upward-propagating waves on a pixel-by-pixel basis between two given heights. To ensure accuracy, we smooth the power, phase difference, and coherency spectra by averaging over a 3 × 3 pixel neighborhood before detection. Our selection process involves choosing pixels where the frequency "\textbf{f}" corresponds to the highest power, maximum coherency, and maximum phase difference, provided that the coherency value exceeds 50$\mu$ (where, $\mu$ represent the mean coherency across the entire FoV and 50 is an arbitary number choosen to study highly coherent waves). We apply this criterion to three central frequencies—1.5, 3.3, and 5.5 mHz—with a bandwidth of ±0.544 mHz. These frequencies represent AGWs (1.5 mHz) and two acoustic wave frequencies (3.3 and 5.5 mHz) observed in the solar atmosphere. Since atmospheric gravity waves exhibit a negative phase difference while transporting energy upward, we adjust our criteria to detect the maximum negative phase difference, keeping the other condition unchanged.

\begin{figure}
	\centering
	\includegraphics[width=85mm]{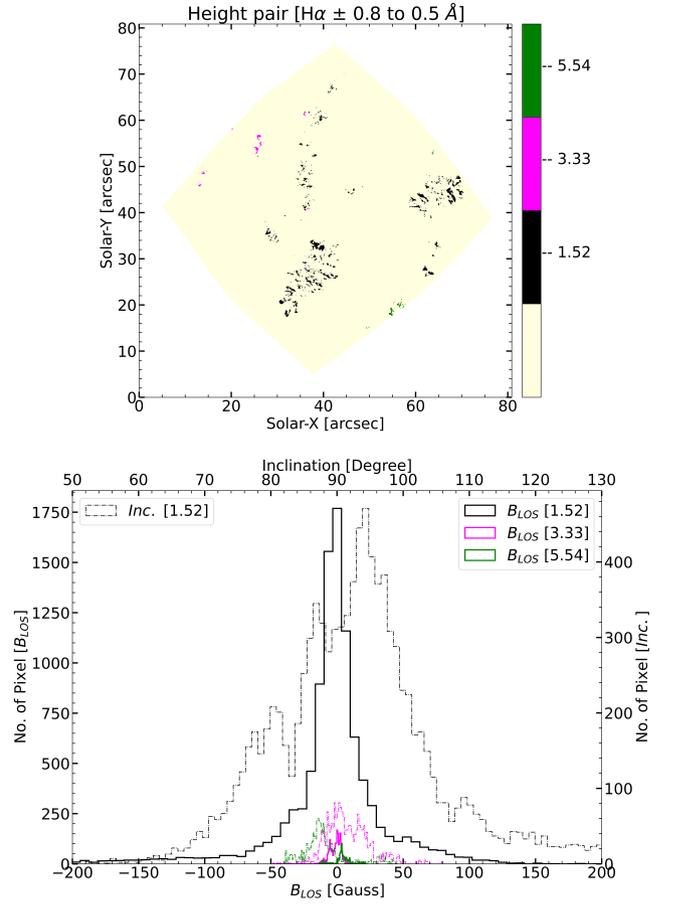}
	\caption{Spatial distribution of the dominant frequencies waves propagating upward as detected at (H$\alpha$ $\pm$ 0.8 \AA~ -- H$\alpha$ $\pm$ 0.5 \AA~) height pair. The left panel shows the spatial distribution of the propagating waves frequencies corresponding to 1.5, 3.3, and 5.5 mHz. The right panel shows the histogram of the number of detected pixels for each frequency. To clearly observe the variations in the histograms corresponding to the 3.3 mHz and 5.5 mHz frequency bands, the number of pixels associated with these frequencies has been increased by a factor of 3.}
	\label{Fig7}
\end{figure}

\begin{figure}
	\centering
	\includegraphics[width=85mm]{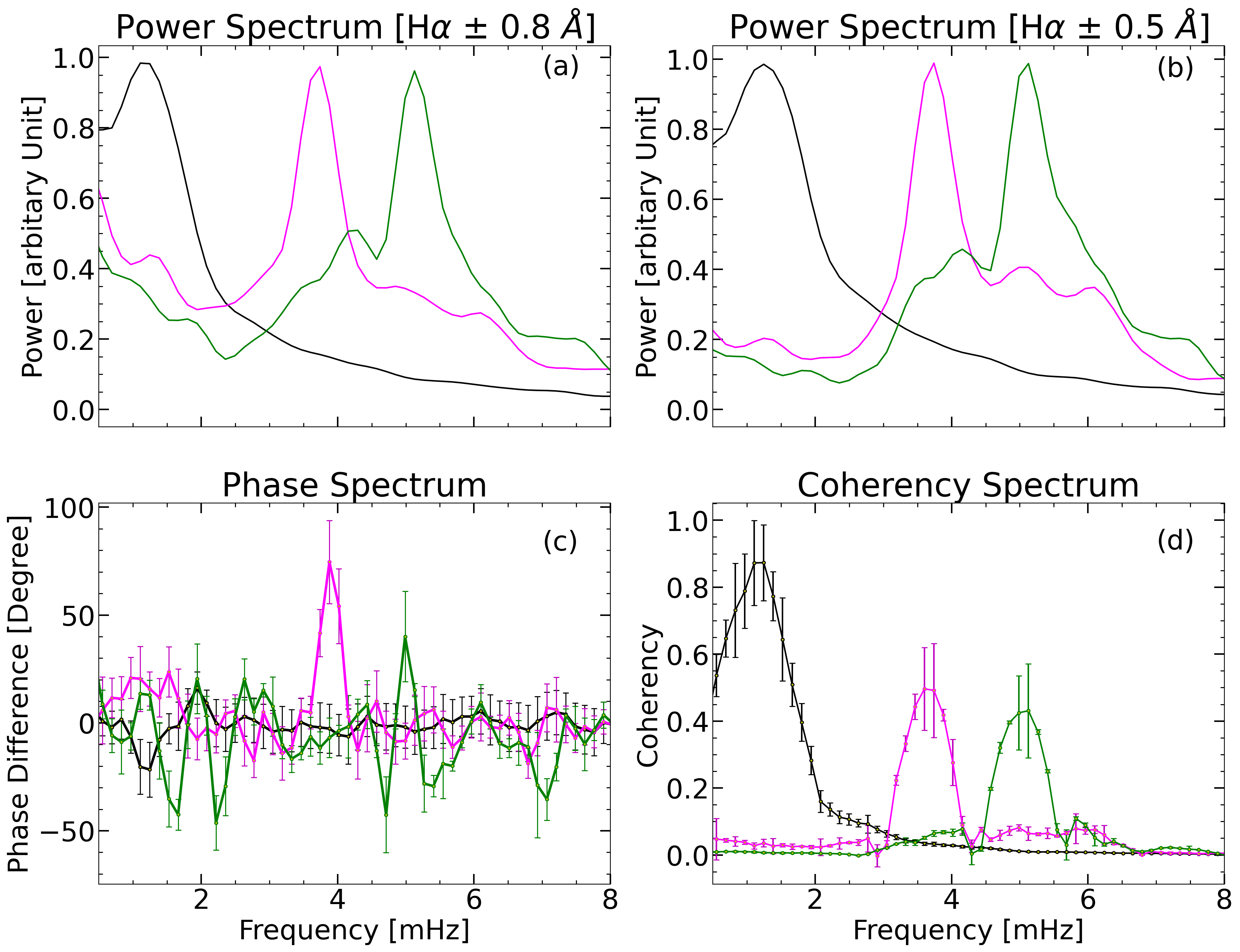}
	\caption{Average power spectrum of the detected pixels in the three frequencies bands near photospheric layer and mid chromospheric layer of the solar atmopshere are shown in panel (a) and panel (b). Panel (c) and panel (d) shows the average phase difference and coherency spectrum of the detected pixels between these two heights. The black, magnenta and green curves are corresponding to the 1.5, 3.3, and 5.5 mHz frequency waves. The vertical lines over the average phase difference and the cohenrency spectrum are the 1 $\sigma$ error bar from their mean position. The strong negative phase difference, high coherency and high power at two heights corresponding to 1.1 mHz frequency are typical signatures of upward-propagating AGWs in the solar atmosphere.}
	\label{Fig8}
\end{figure}

Initially, we focus on detecting propagating waves by examining the solar atmospheric layers represented by H$\alpha$ $\pm$ 0.8 \AA~ (near-photospheric) and H$\alpha$ $\pm$ 0.5 \AA~ (mid chromospheric). After applying the threshold criteria within the three frequency bands, we present the distribution of detected pixel locations as a 2D map in the left panel of Figure \ref{Fig7}. The right panel of Figure \ref{Fig7} displays the histogram of LOS magnetic field and their inclination of these detected pixels for different frequency bands. Panels (a) and (b) of Figure \ref{Fig8} present the average power spectrum of the detected pixels in the near-photospheric and mid chromospheric layers, respectively. Panels (c) and (d) illustrate the phase difference and coherency spectra between these two atmospheric heights. Comparisons of the average power, phase difference, and coherency spectra reveal that the low-frequency band shows significant power, coherency, and negative phase difference at approximately 1.1 mHz—characteristic signatures of upward-propagating AGWs. Conversely, the other two frequency bands exhibit positive phase differences (though less pronounced for the 3.3 mHz band), along with significant power and coherency, indicating their upward propagation as well. The 2D map of detected propagating waves shows that AGWs (indicated in black) are concentrated in or mostly around regions of magnetic flux concentration. In contrast, mid and high-frequency waves (magenta and green) are generally located away from these regions. Additionally, the number of pixels associated with AGWs is notably higher than those for the other two frequency bands.

\begin{figure}
	\centering
	\includegraphics[width=85mm]{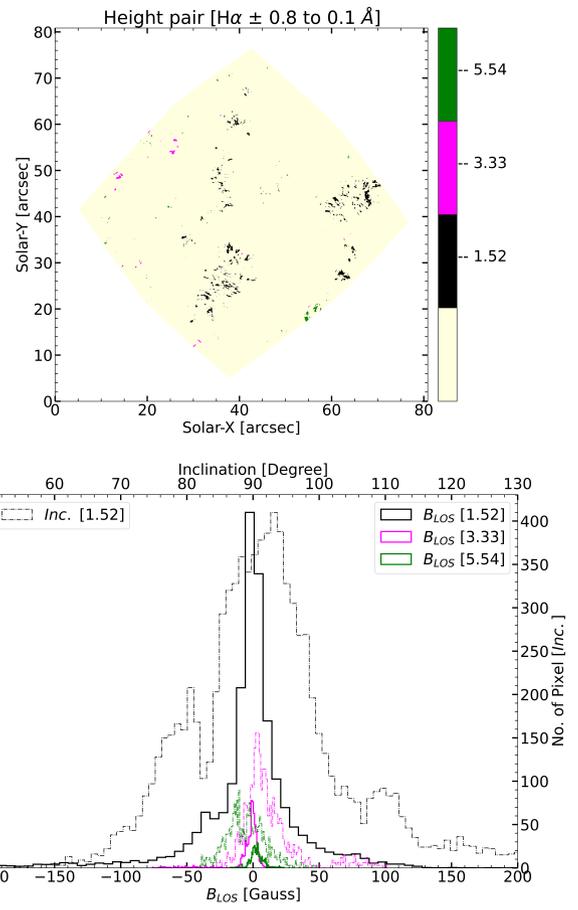}
	\caption{Same as Figure \ref{Fig7} but for the near photosphere (H$\alpha$ $\pm$ 0.8 \AA) and upper chromosphere (H$\alpha$ $\pm$ 0.1 \AA) height pair.}
	\label{Fig9}
\end{figure}

\begin{figure}
	\centering
	\includegraphics[width=85mm]{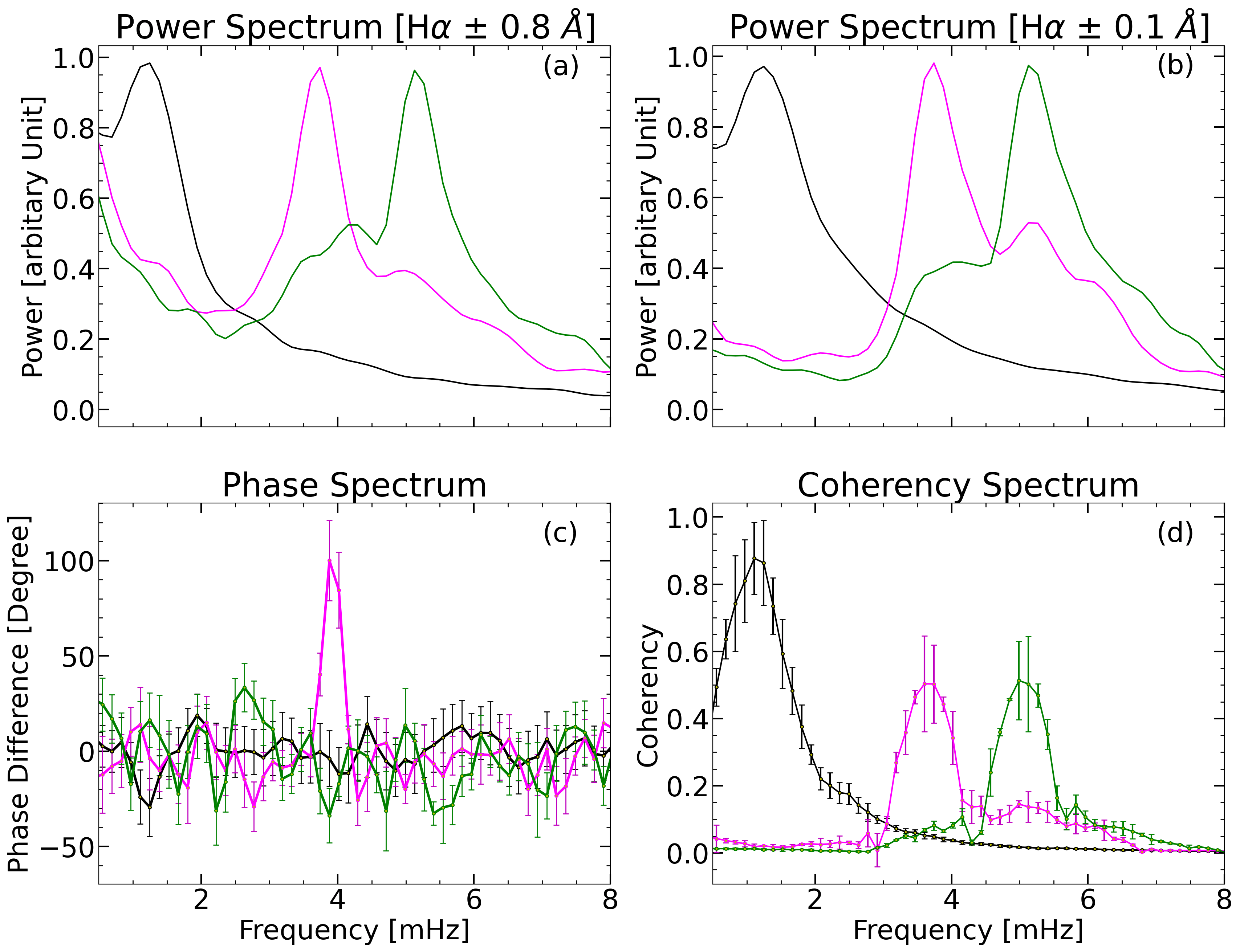}
	\caption{Same as Figure \ref{Fig8} but for the near photosphere (H$\alpha$ $\pm$ 0.8 \AA) and upper chromosphere (H$\alpha$ $\pm$ 0.1 \AA) height pair. }
	\label{Fig10}
\end{figure}

To further validate the tracing of these waves originating near the photosphere and reaching the upper atmosphere, we repeated the analysis for height-pair 
(H$\alpha$ $\pm$ 0.8 \AA~--H$\alpha$ $\pm$ 0.1 \AA), which corresponds to larger height difference than the previous one in the solar atmosphere. The spatial distribution of detected pixels across the 1.5, 3.3, and 5.5 mHz frequency bands is illustrated in Figure \ref{Fig9}, along with a histogram of LOS magnetic field and their inclination. Figure \ref{Fig10} provides the average power, phase difference, and coherency spectra for these detected pixels. Finally, to confirm that these waves indeed travel from the photosphere through the chromosphere to higher chromospheric layers, we apply the same selection criteria for Fe I 6173 \AA -- H$\alpha$ $\pm$ 0.1 \AA~ height pair.

\begin{figure}
	\centering
	\includegraphics[width=85mm]{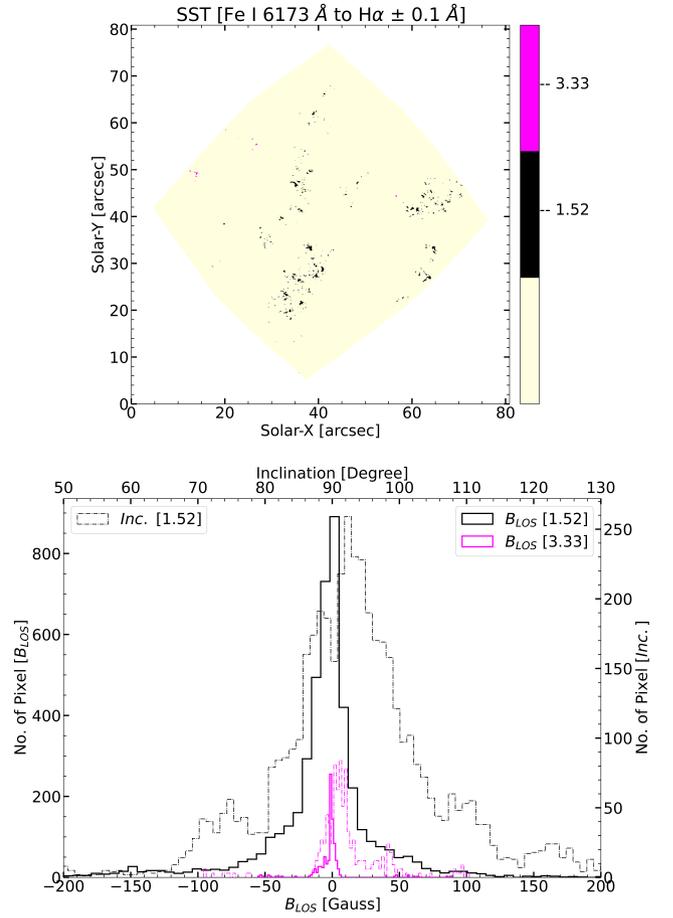}
	\caption{Same as Figure \ref{Fig7} but for the photosphere (Fe I 6173 \AA) and upper chromosphere (H$\alpha$ $\pm$ 0.1 \AA) height pair.}
	\label{Fig11}
\end{figure}

\begin{figure}
	\centering
	\includegraphics[width=85mm]{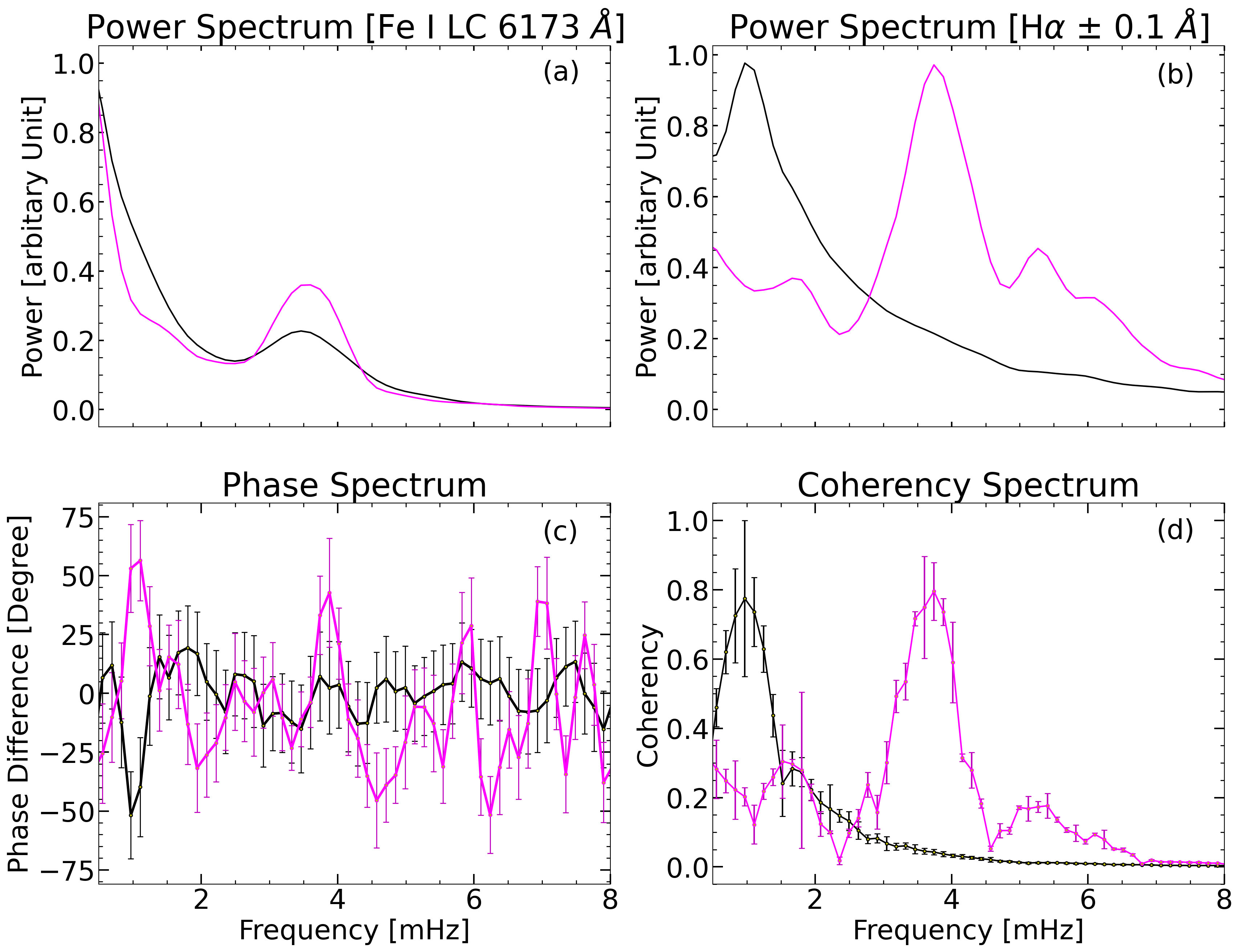}
	\caption{Same as Figure \ref{Fig8} but for the photosphere (Fe I 6173 \AA) and upper chromosphere (H$\alpha$ $\pm$ 0.1 \AA) height pair. }
	\label{Fig12}
\end{figure}

Here, we use average bisector velocity as a time-varying signal to determine the power spectrum, phase difference and coherency spectra with other atmospheric layers for photospheric spectral line as discussed in Section \ref{ssec3}. Figures  \ref{Fig11} and \ref{Fig12} show the results obtained from the similar analysis as mentioned for the previous two height pairs. Although the photosphere do not show enhanced power at 1.1 mHz, the presence of enhanced power at chromopsheric height along with negative phase difference indicates the propagation of AGWs. On the other hand, we do not found any pixels associated with propagating waves at the 5.5 mHz frequency. It should be noted that in order to observe the variations in the histograms corresponding to the 3.3 mHz and 5.5 mHz frequency bands, the number of pixels associated with these frequencies has been increased by a factor of 3 in each case of H$\alpha$.

To further support the propagation of these waves in the solar atmosphere, we also analyzed the multi-height velocity pairs for Ca II IR observations. Following the approach used for H$\alpha$, we initially selected velocity pairs at Ca II IR $\pm$ 0.4 \AA~ (lower chromosphere) and Ca II IR $\pm$ 0.225 \AA~ (below mid-chromosphere), applying the same detection criteria. The results are shown in Figure \ref{FigA5} and Figure \ref{FigA6}. Next, we replaced the below mid-atmosphere with Ca II IR $\pm$ 0.075 \AA~ and applied the criteria again, with results presented in Figure \ref{FigA7} and Figure \ref{FigA8}. Finally, to determine if the waves originate from the photosphere and reach the mid-chromosphere, we used the photosphere (Fe I 6173 \AA) for the lower atmosphere and Ca II IR $\pm$ 0.075 \AA~ for the upper atmosphere. The results for this height pair are shown in Figure \ref{FigA9} and Figure \ref{FigA10}. The findings from Ca II IR observations are consistent with those obtained from H$\alpha$  and hence reinforcing the conclusions drawn about wave propagation in the solar atmosphere. 

For completeness, we also used the height pair (H$\alpha$ $\pm$ 0.8 Å -- Ca II IR $\pm$ 0.075 Å), which indicates near photosphere -- mid chromosphere height, and applied our threshold criteria. The spatial distribution of the detected pixels and the histogram of the magnetic field strength and inclination are shown in Figure \ref{FigA11} and Figure \ref{FigA12}, respectively. The results are similar to the results obtained with the height pair (H$\alpha$ $\pm$ 0.8 \AA -- H$\alpha$ $\pm$ 0.1 \AA).

In addition to the aforementioned method for studying propagating waves in the solar atmosphere, we have also applied the phase difference spectra technique for two height pairs: H$\alpha \pm 0.8$ \AA\ -- H$\alpha \pm 0.5$ \AA\ and H$\alpha \pm 0.8$ \AA\ -- H$\alpha \pm 0.1$ \AA. This method involves estimating the 3D Fourier transform of the time-varying signals at two heights, followed by azimuthal averaging in the $k_x - k_y$ plane to generate the horizontal wavenumber–frequency ($k_h - \nu$) diagram (\cite{2008ApJ...681L.125S,2017ApJ...835..148V,2023AdSpR..72.1898K,2023ApJ...952...58V}). 
The $k_h - \nu$ diagram for the V--V signals corresponding to the H$\alpha \pm 0.8$ \AA\ -- H$\alpha \pm 0.5$ \AA\ height pair is presented in Figure~\ref{k_w_diagram}. A distinct blue patch, indicating a negative phase difference, is observed at low frequencies and low wavenumbers, which is a signature of AGWs in the solar atmosphere. Similar results were also obtained for H$\alpha \pm 0.8$ \AA\ -- H$\alpha \pm 0.1$ \AA\ height pair with increased phase difference. However, $k_h - \nu$ diagram do not provide us the information related to location of the propagating waves with a particular frequency. Hence, we prefer the earlier approach for the study of propagating waves in the solar atmosphere at different regions, viz., magnetic and non-magnetic regions.

\begin{figure}
	\centering
	\includegraphics[width=85mm]{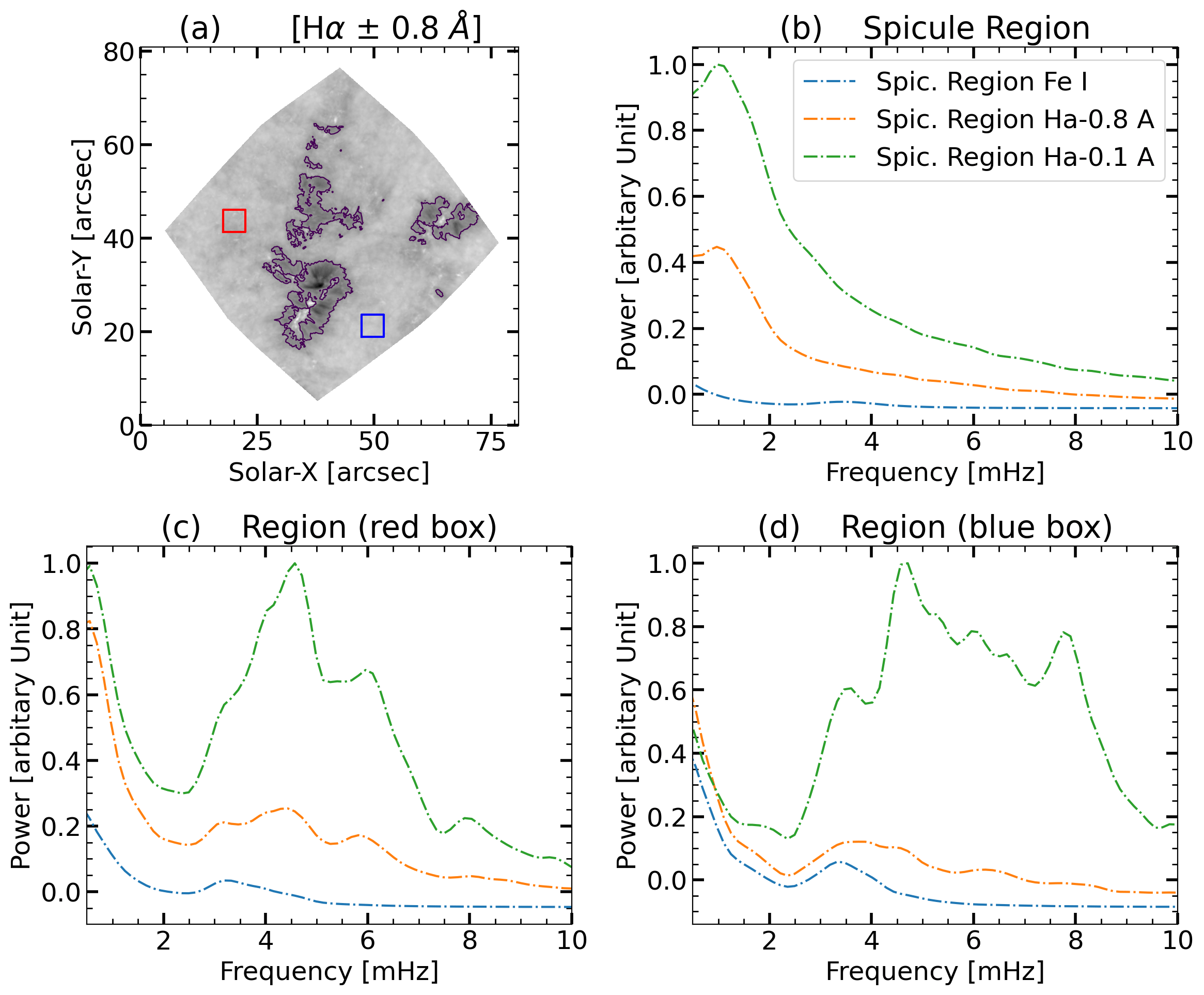}
	\caption{Panel \textbf{(a)} depicts the average wing filtergram at H$\alpha$ $\pm$ 0.8 \AA. The black contour over it represents the location of spicules. The two colored boxes (red and blue) served as the QS and are deviod of the spicules. Panel (b), (c) and (d) are the average power spectrum of spicular locations, and the two QS region (red and blue) at H$\alpha$ $\pm$ 0.8 \AA, H$\alpha$ $\pm$ 0.5 \AA~ and  H$\alpha$ $\pm$ 0.1 \AA. The comparison of the power spectrum of the spicular and non spicular location suggest that spicular region exhibit oscillation corresponding to AGWs frequency. }
	\label{Fig13}
\end{figure}

As both spicules and propagating AGWs are associated with magnetic flux concentration regions, we investigate the relationship between the oscillations associated with the spicules and the propgating AGWs. In this regard, we examined the average power spectrum corresponding to the average location of the spicules at different heights in the solar atmosphere. Additionally, we compared the oscillations associated with spicules to those in QS regions, which are free of spicules. Panel (a) of Figure \ref{Fig13} displays the average wing filtergram of H$\alpha$ (H$\alpha$ $\pm$ 0.8 Å) with contours at 97 \% of the mean intensity, highlighting the locations of spicules. The red and blue boxes represent QS regions without spicules, which were used to calculate the average power spectra for these spicule-free areas. Panel (b) presents the average power spectrum for the spicular regions, excluding the areas where spicules overlap with magnetic flux concentration regions. Panels (c) and (d) display the average power spectra for the QS regions outlined by the red and blue boxes, respectively.

\section{Summary and Discussion}  \label{Sec4}

Atmospheric gravity waves (AGWs) are known for transporting energy from the photosphere to chromosphere, exhibiting a characteristic negative phase difference \citep{2008ApJ...681L.125S}. The estimated energy flux of AGWs is comparable to the radiative losses of the chromosphere \citep{2008ApJ...681L.125S,2011A&A...532A.111K} together with impact of magnetic field on their propagation makes AGWs an important aspect to study in the chromopshere. Studies by \cite{2010MNRAS.402..386N,2011MNRAS.417.1162N,2016ApJ...828...88H} and \cite{2017ApJ...835..148V,2023AdSpR..72.1898K} suggests that AGWs are either absent or partially reflected back into lower atmospheric layers due to the influence of vertical magnetic field topology. Conversely, horizontal magnetic fields facilitate the waves propagation to chromospheric heights. While the impacts of vertical and horizontal magnetic fields on AGWs have been studied, the effect of inclined magnetic fields on AGW propagation, particularly in the high chromosphere, remains underexplored.

In order to understand the propagation of AGWs and the acoustic waves with frequencies centered at 3.3 and 5.5 mHz, we analyze and compare the power distribution at various heights in the solar atmosphere using velocities obtained employing filtergram difference and bisector velocity as the time-varying signals.

Despite the fact that velocity obtained from filtergram difference and bisector method showed siginificant spread, power maps obtained by both methods revealed almost similar variations in power distribution at all heights in the solar chromosphere. Also, the dominant power corresponding to certain frequencies remains the same (refer to Figures \ref{Fig6} and Figure \ref{FigA3}). This consistency is evident in Figure \ref{Fig3}, where the variation pattern of the average signal originating from certain heights in the solar atmosphere is almost identical, irrespective of whether it is derived from filtergram difference or Doppler velocity. This indicates that the nature of the plasma flow motion (upflow or downflow) at any given location remains same and is independent of the method used.

Despite the different line formation mechanisms of the two chromospheric lines and the fact that the studied chromospheric layers sample different parts of the atmosphere (e.g., the atmospheric layer corresponding to H$\alpha$ $\pm$ 0.8 \AA~ and Ca II $\pm$ 0.6 \AA~ are not the same), their variation patterns show a good match. Furthermore, the variation pattern of the average signal at different heights in the solar atmosphere also looks similar, suggesting that different atmospheric layers could be oscillating at the same frequencies. This is consistent with the results shown in Figure \ref{Fig6} and Figure \ref{FigA3}, where dominant powers corresponding to the same frequencies are evident across various atmospheric layers.
This dominance is also clearly seen in Figures \ref{Fig4} and \ref{Fig5}, where the magnetic regions of the solar atmosphere consistently show enhanced power in each layer within the AGW frequency regime. This observation aligns well with the propagation characteristics of AGWs, which have a vertical wavelength of a few thousand kilometers in the upper photosphere/lower chromosphere (\cite{1981ApJ...249..349M}), causing the atmosphere to oscillate in sync with the AGWs frequencies. Therefore, to further study the propagating waves in the solar atmosphere, we estimate the phase difference and coherency spectrum between different height pairs. These propagating waves are averaged and studied in the same frequency bands for which the power map at different heights is averaged.

To identify the spatial locations and study the propagation of waves in the solar atmosphere, we perform a pixel-by-pixel analysis by applying threshold criteria to the power, phase difference, and coherency spectra for three height pairs: H$\alpha$ $\pm$ 0.8 Å -- H$\alpha$ $\pm$ 0.5 Å, followed by H$\alpha$ $\pm$ 0.8 Å -- H$\alpha$ $\pm$ 0.1 Å, and then Fe I 6173 Å -- H$\alpha$ $\pm$ 0.1 Å. Similarly, for Ca II IR, we examine height pairs such as Ca II IR $\pm$ 0.4 Å -- Ca II IR $\pm$ 0.225 Å, Ca II IR $\pm$ 0.4 Å -- Ca II IR $\pm$ 0.075 Å, and Fe I 6173 Å -- Ca II IR $\pm$ 0.075 Å. The height pairs are chosen so that the height difference increases progressively from one pair to the next for a given H$\alpha$ or Ca II IR spectrum. In addition, we have also considered a height pair ( H$\alpha$ $\pm$ 0.8 Å -- Ca II IR $\pm$ 0.075 \AA) for cross verification in two chromospheric spectral lines and found that the results are similar to the results obtained with the spectral lines H$\alpha$ and Ca II IR individually.

The spatial distribution of pixels which exhibit characteristics of AGWs are concentrated in or near magnetic flux concentration regions. The average power, phase difference, and coherency spectra of the detected pixels characterized by a distinctive negative phase difference around the 1.1 mHz frequency and the phase difference increases with the increase in height difference, indicating upward wave propagation of AGWs. The histogram of LOS magnetic field strength and inclination for AGW pixels reveals that most of them are associated with near-zero LOS magnetic fields and nearly horizontal magnetic field inclinations. These pixels are typically found in close proximity to, but not within, magnetic flux concentration regions. Although QS locations also exhibit similar characteristics (near-zero magnetic fields and nearly horizontal inclinations), the key difference for AGWs is their abundance near the magnetic flux concentration regions rather than away from them. Additionally, a significant number of pixels correspond to AGWs are associated with LOS magnetic fields greater than ±50 G with notable inclinations. Similar analysis was performed for Ca II IR height pairs, yielding nearly identical results to those from H$\alpha$. 
Due to the presence of AGWs in or near-magnetic flux concentration regions, it is speculated that the magnetic inclination might plays a role in the upward propagation of AGWs.

It is important to note that the power spectrum of pixels associated with AGWs also exhibited significant power around the 3.3 mHz frequency in the photosphere, a typical oscillation frequency of the QS photosphere, when analyzing the height pair Fe I 6173 Å -- H$\alpha$ $\pm$ 0.1 Å. This power at 3.3 mHz in the photosphere was expected due to the weak magnetic field in the observed region. However, these pixels did not displayed significant power at the 3.3 mHz frequency in higher atmospheric layers but also lacked notable phase differences and coherency, suggesting that the wave is non-propagating and confined to the photosphere in the pixels corresponding to AGW frequencies.

In contrast, for acoustic regimes, the majority of detected pixels are found in QS region. The oscillatory power in acoustic regime also exhibit increase in power and coherency, with a finite phase difference suggesting upward propagation, but they are much fewer in number compared to pixels exhibiting characteristics of AGWs. On the other hand, the analysis of power maps corresponding to the height pair Fe I 6173 Å -- H$\alpha$ $\pm$ 0.1 Å and Fe I 6173 \AA~ -- Ca II IR $\pm$ 0.075 \AA~ reveals no pixels  corresponding to 5.5 mHz frequency band, although the spatial locations for AGWs and the 3.3 mHz frequency remained consistent in this height pair. The absence of 5.5 mHz pixels suggests that these waves do not originate from the photosphere or below but rather from the chromosphere, at least in QS regions. 

Throughout the analysis, it becomes clear that the spatial locations of pixels corresponding to AGW frequencies are associated with or near magnetic flux concentration regions. This is consistent with the known origin of spicules, which are also observed to form in or near these regions \citep{2004Natur.430..536D, 2009ApJ...705..272R,2019Sci...366..890S,2024ApJ...970..179C}. To explore the potential link between spicules and AGWs, we investigated the oscillations at the average locations of spicules and found that these pixels exhibit clear oscillations at multiple heights in the solar atmosphere, corresponding to AGW frequencies. We speculate that these low-frequency AGWs propagate upward into the chromosphere, following the spicular structures, with the spicules acting as waveguides. In contrast, oscillations in non-spicular or quiet Sun regions do not show the presence of AGW frequencies. Although spicules are known to exhibit high-frequency oscillations \citep{2017NatSR...743147S, 2018ApJ...854....9S, 2021ApJ...921...30S, 2022ApJ...930..129B}, we hypothesize that the low-frequency AGWs were not detected in previous studies using space-time map due to the limited lifetime of spicules. This suggests that these low-frequency oscillations associated with AGWs are distinct from the higher-frequency oscillations typically observed in spicules, indicating different dynamics and phenomena in the solar atmosphere.

\section{Conclusions} \label{Sec5}

We analyzed and compared the power distribution at various heights in the solar atmosphere utilising velocity obtained using filtergram difference and bisector methods as time-varying signals, focusing on frequencies corresponding to AGWs and acoustic regimes, using H$\alpha$ and Ca II spectra. The results revealed a consistent, regular increase in power with height, and the same dominant power frequencies were observed at any given pixel, despite the significant spread in velocities between the two methods. In addition, we identified the spatial locations of propagating waves in the same frequency bands, finding that AGWs propagate upward into the solar chromosphere after being detected in or near the photosphere. Our results suggest that AGWs are originated at the location of spicules and propagate upward along the spicular body. Although we did not estimate the energy budgets of these waves, the observed power distribution at different heights indicates that AGWs may carry enough energy to compensate for the radiative losses in the magnetic regions in the solar chromosphere. Estimating these energy budgets and radiative losses will be the focus of future work.

\section*{Acknowledgements}

We sincerely thank the referee for their invaluable comments and suggestions. The Swedish 1 m Solar Telescope is operated on the island of La Palma by the Institute for Solar Physics of Stockholm University in the Spanish Observatorio
del Roque de los Muchachos of the Instituto de Astrofísica de Canarias. We would like to thank SST team for making the data publicly available.

\section*{Data Availability}

The observational data used in this study was obtained from the Swedish Solar Telescope (SST) and is available online at the following link: \url{https://dubshen.astro.su.se/sst_archive/search}. \newline



\bibliographystyle{mnras}
\bibliography{Tracing_AGWs} 




\appendix 	
\label{appexA}

\section{Results obtained for Ca II IR Observation}

\begin{figure*}
	\centering
	\includegraphics[width=157mm]{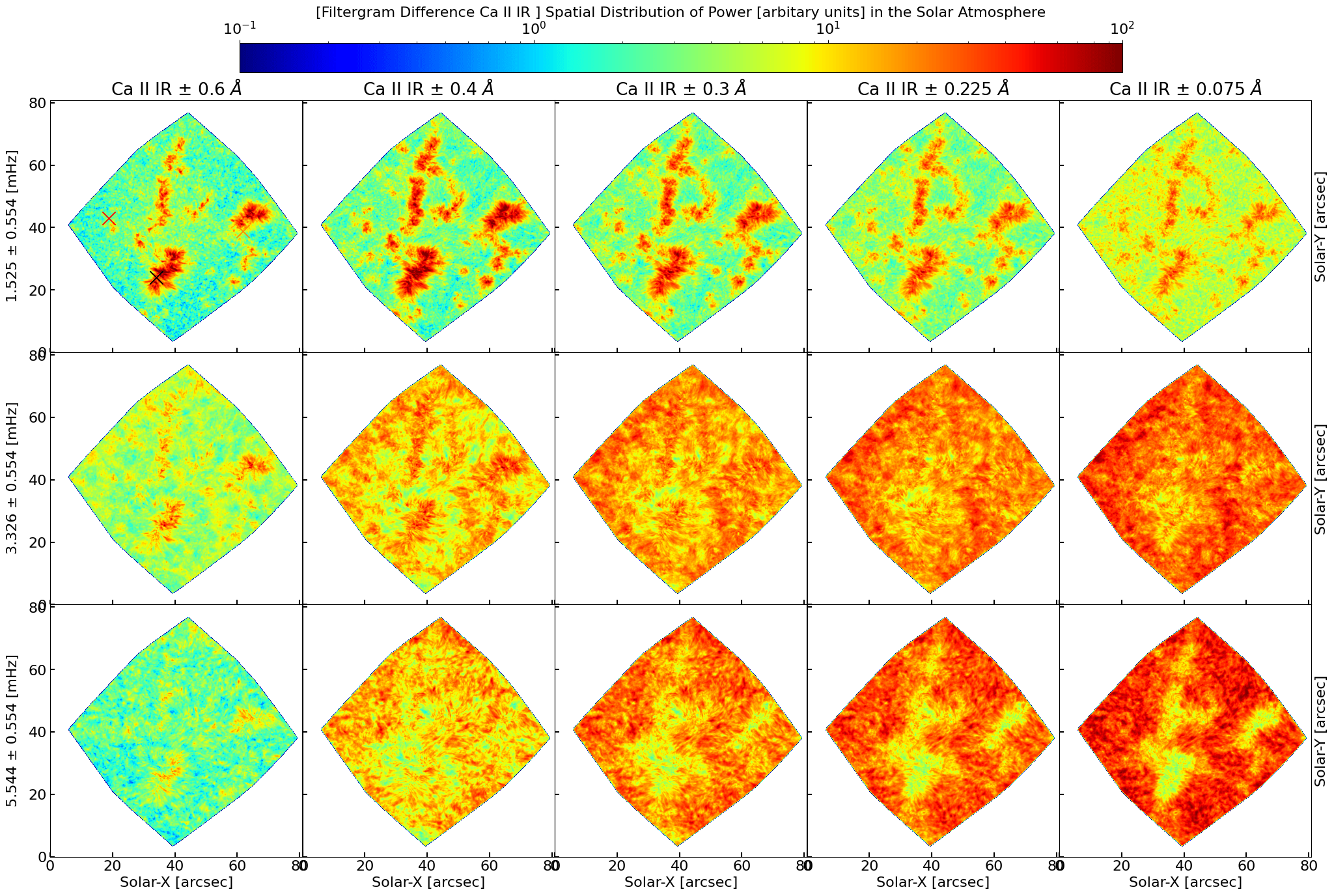}
	\caption{Same as Figure \ref{Fig4} but for Ca II IR. The power spectrum corresponding to the three different locations ( \textbf{X} marks, as shown in the upper right panel) is depicted in the upper panel of Figure \ref{FigA3}.}
	\label{FigA1}
\end{figure*}

\begin{figure*}
	\centering
	\includegraphics[width=157mm]{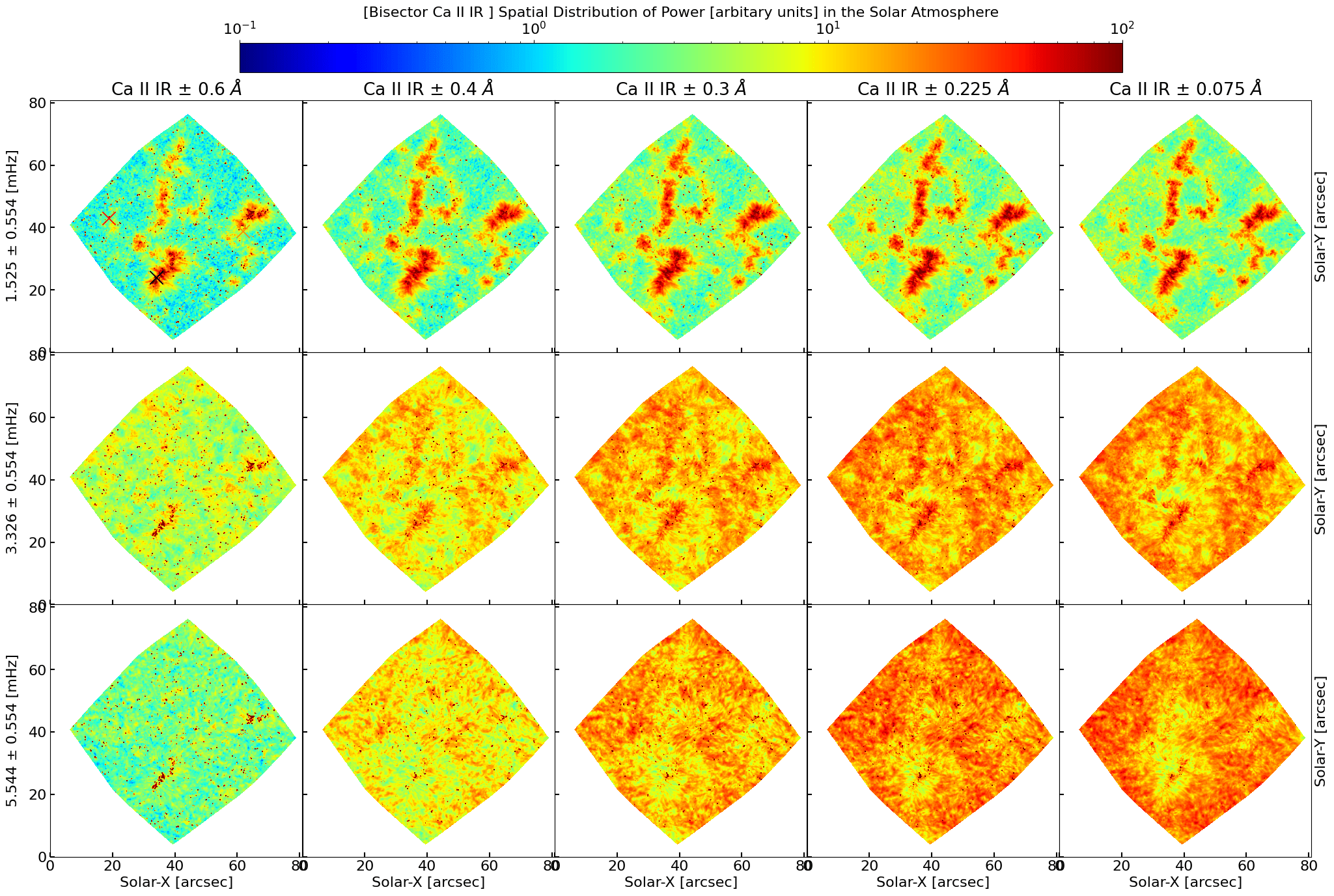}
	\caption{Same as Figure \ref{Fig5} but for the Doppler velocity of Ca II IR. The power spectrum corresponding to the three different locations ( \textbf{X} marks, as shown in the upper right panel) is depicted in the lower panel of Figure \ref{FigA3}.}
	\label{FigA2}
\end{figure*}

\begin{figure}
	\centering
	\includegraphics[width=85mm]{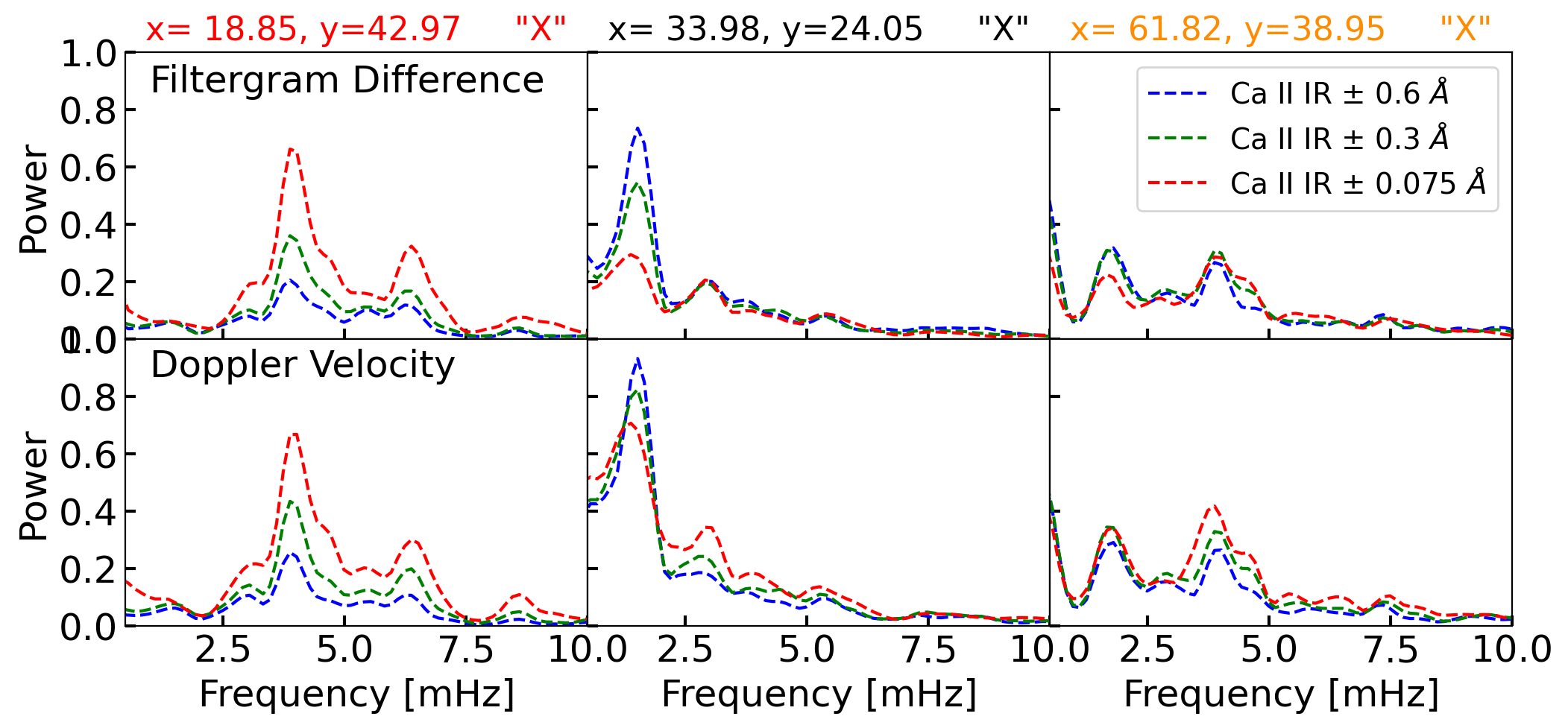}
	\caption{Same as Figure \ref{Fig6} but for Ca II IR showing three power spectra (corresponding to three spatial locations marked by colored "\textbf{X}" in Figure  \ref{FigA1} and Figure \ref{FigA2} from left to right) obtained at three heights in the solar atmosphere using filtergram difference (upper panel) and bisector velocity (lower panel) as a time-varying signal.}
	\label{FigA3}
\end{figure}

\begin{figure}
	\centering
	\includegraphics[width=85mm]{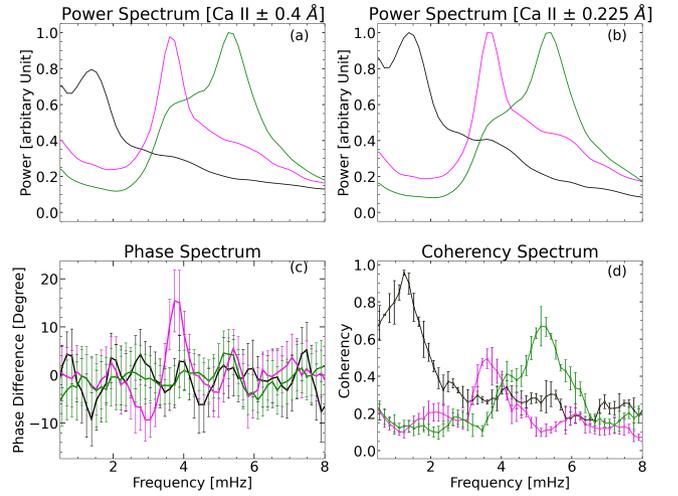}
	\caption{Similar to \ref{Fig7} but for Ca II IR $\pm$ 0.4 \AA~ and Ca II IR $\pm$ 0.225 \AA~ height pair.}
	\label{FigA5}
\end{figure}

\begin{figure}
	\centering
	\includegraphics[width=85mm]{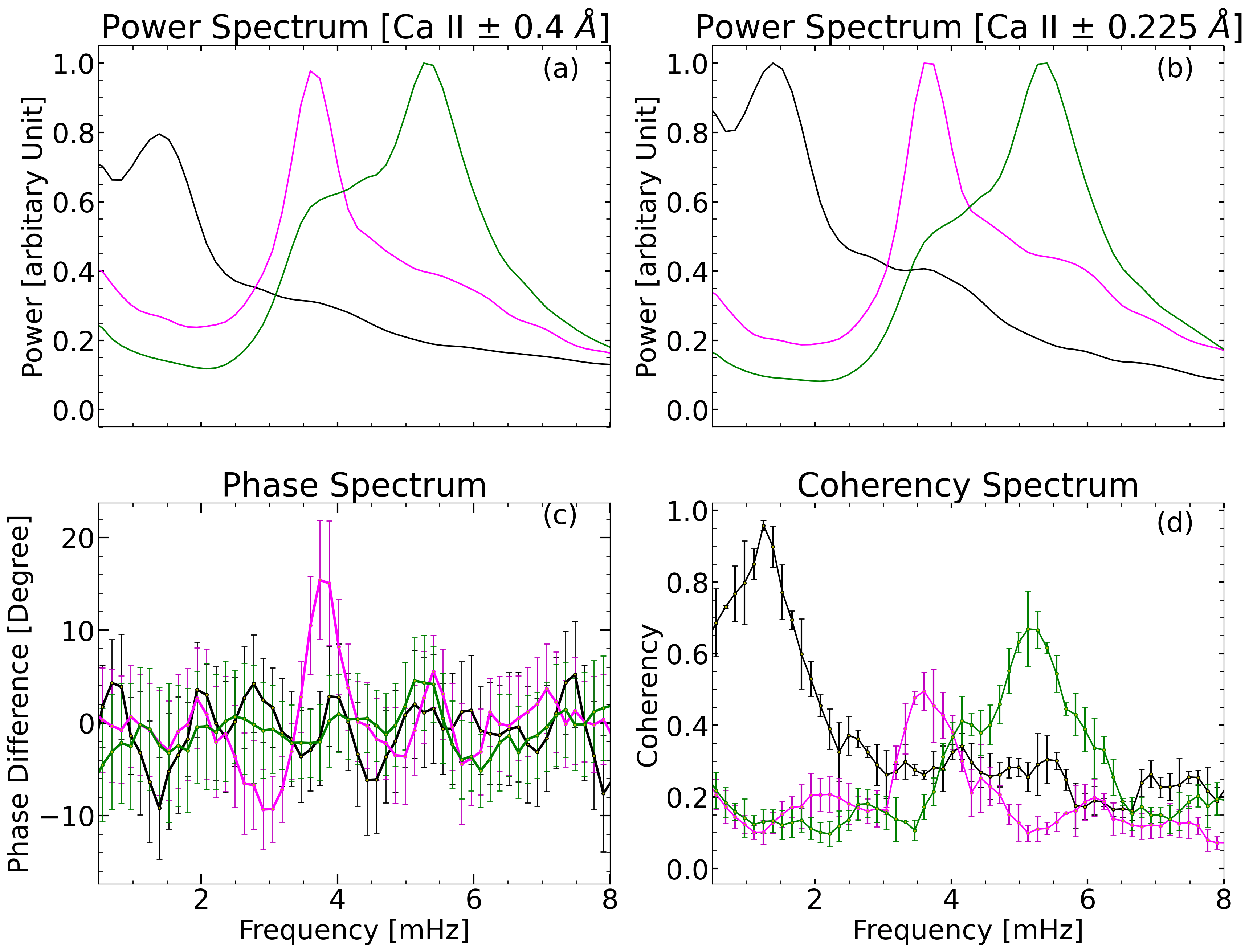}
	\caption{Similar to \ref{Fig8} but for Ca II IR $\pm$ 0.4 \AA~ and Ca II IR $\pm$ 0.225 \AA~ height pair.}
	\label{FigA6}
\end{figure}

This section presents the results obtained for the Ca II IR spectral line (Figures \ref{FigA1} to \ref{FigA12}), derived using an analysis similar to that applied to the H$\alpha$ spectral line as discussed in Section \ref{Sec3} and Section \ref{Sec4}.

\begin{figure}
	\centering
	\includegraphics[width=85mm]{figures/FigA7.png}
	\caption{Similar to \ref{Fig9} but for Ca II IR $\pm$ 0.4 \AA~ and Ca II IR $\pm$ 0.075 \AA~ height pair. }
	\label{FigA7}
\end{figure}

\begin{figure}
	\centering
	\includegraphics[width=85mm]{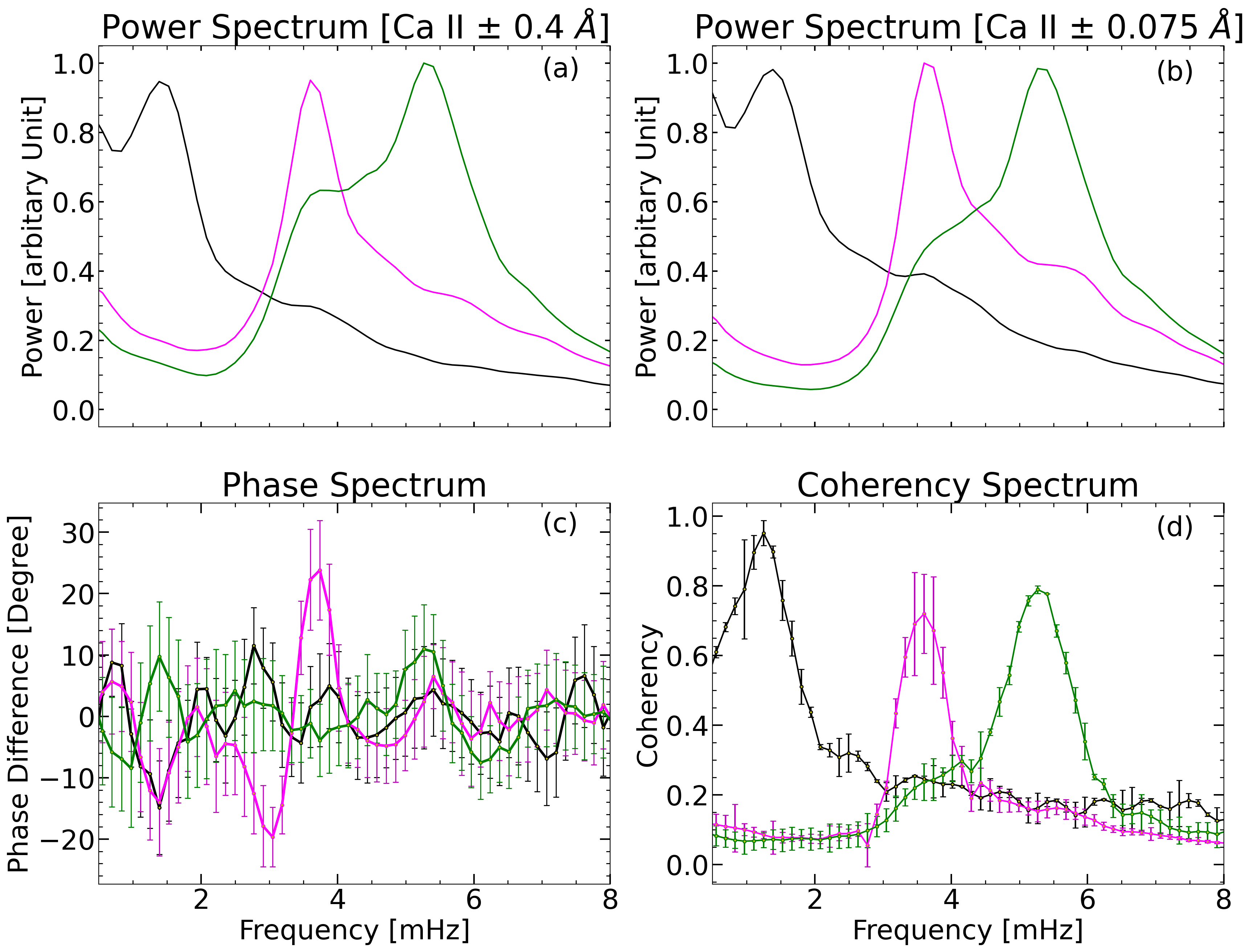}
	\caption{Similar to \ref{Fig10} but for Ca II IR $\pm$ 0.4 \AA~ and Ca II IR $\pm$ 0.075 \AA~ height pair.}
	\label{FigA8}
\end{figure}

\begin{figure}
	\centering
	\includegraphics[width=85mm]{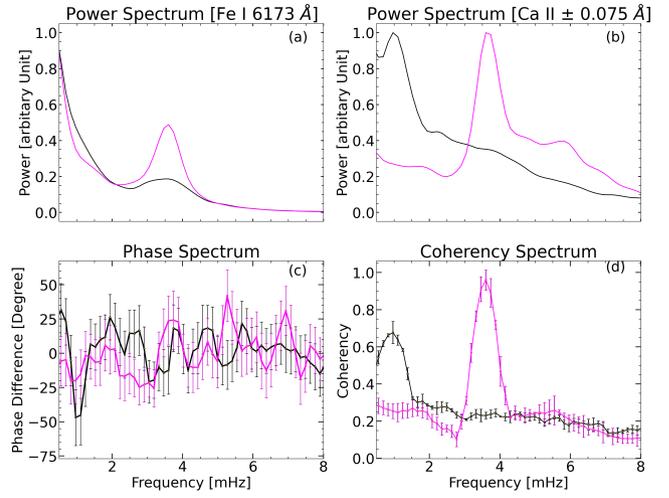}
	\caption{Similar to \ref{Fig11} but for photospheric Fe I 6173 \AA~ and Ca II IR $\pm$ 0.075 \AA~ height pair.}
	\label{FigA9}
\end{figure}

\begin{figure}
	\centering
	\includegraphics[width=85mm]{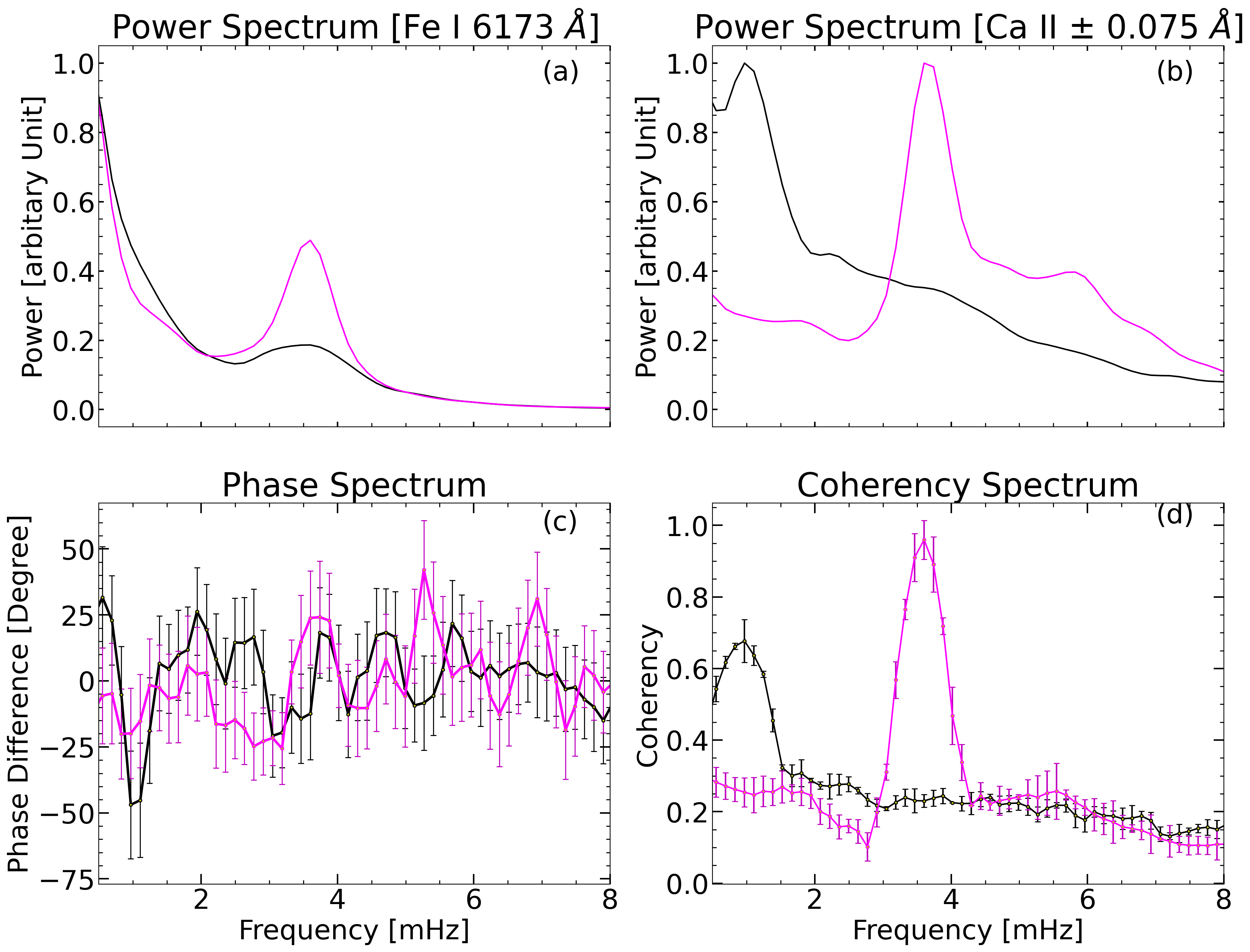}
	\caption{Similar to \ref{Fig12} but for photospheric Fe I 6173 \AA~ and Ca II IR $\pm$ 0.075 \AA~ height pair.}
	\label{FigA10}
\end{figure}

\begin{figure}
	\centering
	\includegraphics[width=85mm]{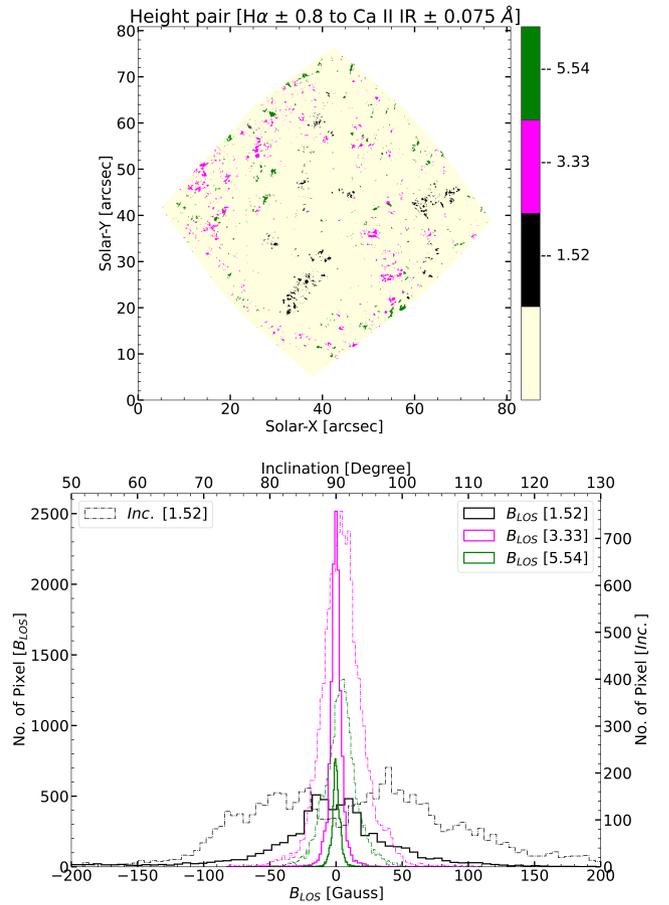}
	\caption{Similar to \ref{Fig7} but for H$\alpha$ $\pm$ 0.8 \AA~ and Ca II IR $\pm$ 0.075 \AA~ height pair.}
	\label{FigA11}
\end{figure}

\begin{figure}
	\centering
	\includegraphics[width=85mm]{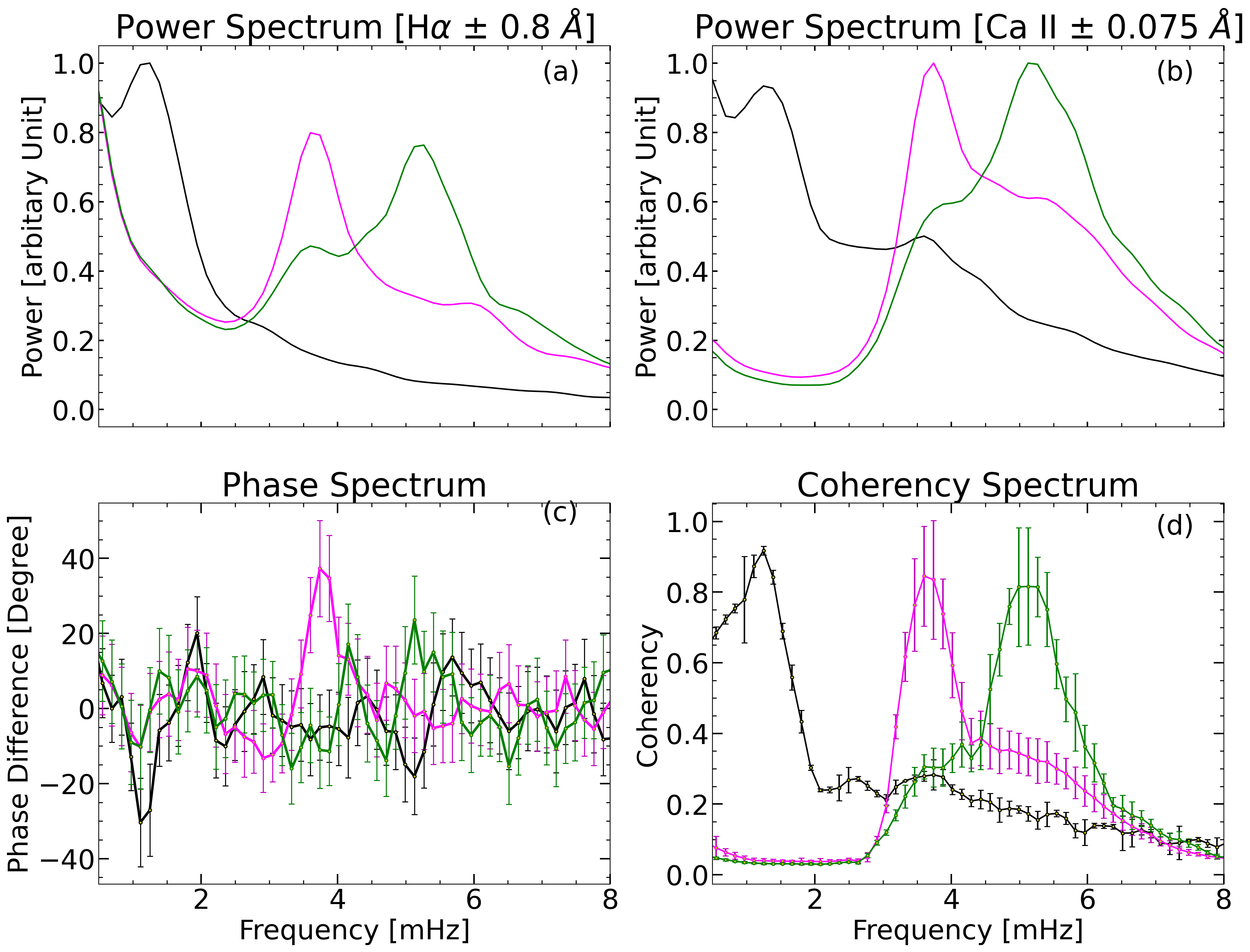}
	\caption{Similar to \ref{Fig8} but for H$\alpha$ $\pm$ 0.8 \AA~ and Ca II IR $\pm$ 0.075 \AA~ height pair.}
	\label{FigA12}
\end{figure}

\section{Phase Difference Spectrum ($k_h - \nu$ diagram)}
\label{appexB}

This section presents the phase difference spectrum ($k_h - \nu$ diagram) corresponding to the H$\alpha$ $\pm$ 0.8 \AA~ -- H$\alpha$ $\pm$ 0.5 \AA~ height pair in the solar atmosphere. The $k_h - \nu$ diagram is constructed by performing a 3D Fourier transform of the time-varying velocity signals (V--V) at two heights, followed by azimuthal averaging in the $k_x - k_y$ plane to derive the horizontal wavenumber–frequency ($k_h - \nu$) diagram. A clear negative phase difference (distinct blue patch) observed at low frequencies and low wavenumbers represents a characteristic signature of AGWs in the solar atmosphere.

\begin{figure}
	\centering
	\includegraphics[width=85mm]{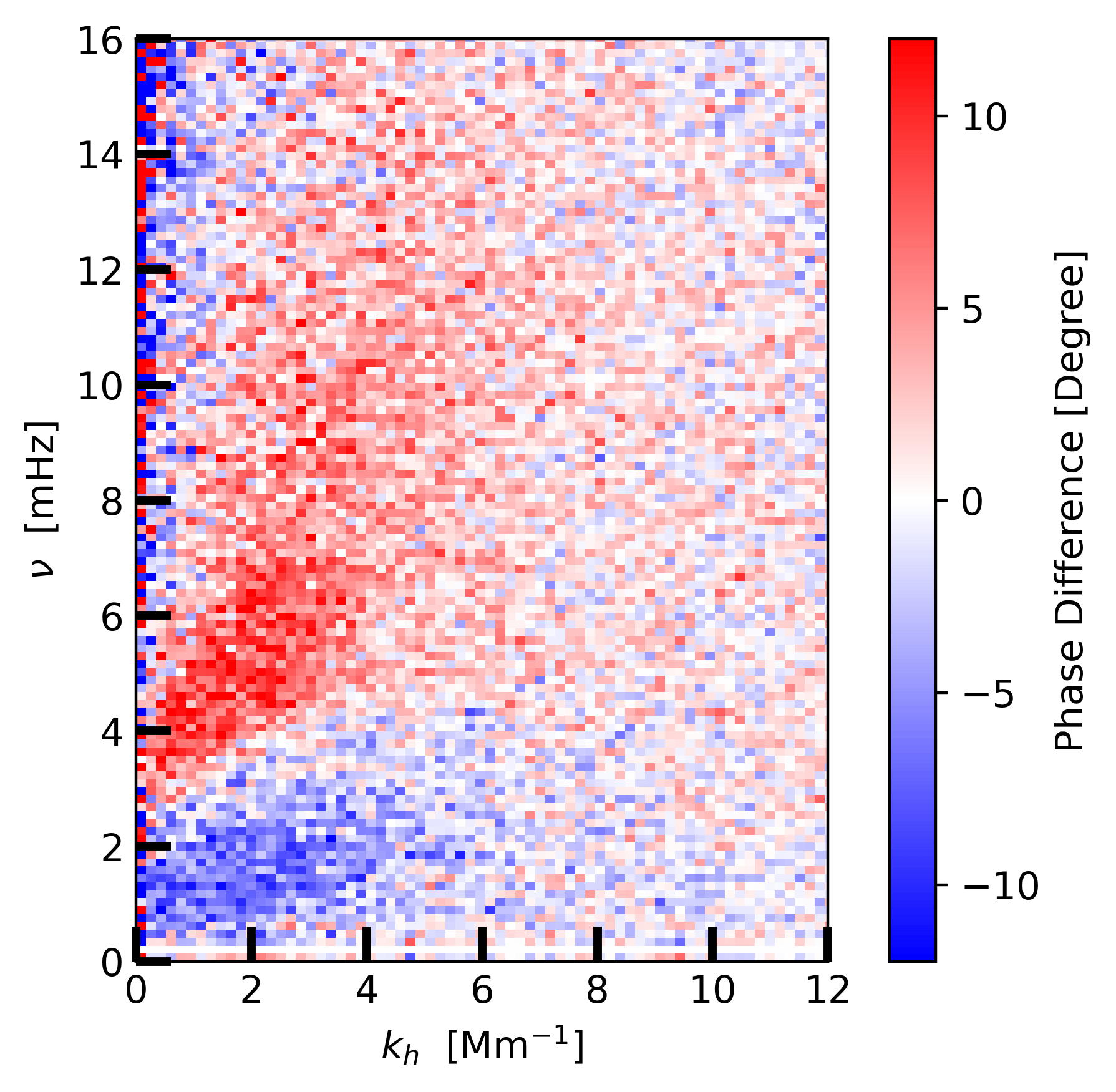}
	\caption{The Figure shows the phase difference specrum for H$\alpha$ $\pm$ 0.8 \AA~ -- H$\alpha$ $\pm$ 0.5 \AA~ height pair.}
	\label{k_w_diagram}
\end{figure}


\bsp	
\label{lastpage}
\end{document}